\documentclass[a4paper,11pt]{article}
\pdfoutput=1 % if your are submitting a pdflatex (i.e. if you have
\usepackage{jcappub} % for details on the use of the package, please
\usepackage[T1]{fontenc} % if needed

\usepackage{comment}
\usepackage{hanging}
\usepackage{orcidlink}

\title{Can a multi-tracer approach improve the constraints on the turnover scale at low redshift?}

\author[a]{Y. Lai,}
\author[a]{C. Howlett,}
\author[a]{T. M. Davis}

\affiliation[a]{School of Mathematics and Physics, The University of Queensland, QLD 4072, Australia.}

\emailAdd{y.lai1@uqconnect.edu.au}

\abstract{The turnover scale of the power spectrum is related to the size of the particle horizon at the matter-radiation equality, which can be used as a standard ruler to constrain cosmological parameters. In this work, we apply a model-independent method to mock datasets to forecast constraints on the turnover scale below a redshift of 0.5, investigating for the first time with a multi-tracer approach. We find that combining the galaxy density with peculiar velocity does not improve the turnover scale constraints for current or currently planned surveys because either the cosmological volume or the effective number density of peculiar velocities is too low. However, we demonstrate that when combining the galaxy power spectrum from 4HS with the HI power spectrum from SKA1-B2, the constraints on the turnover scale improve by \(\sim30\%\) compared to using only a single tracer. We demonstrate for the first time that combining DESI, 4HS, and an SKA Phase 1 Band 2 survey could achieve a \(\sim5\%\) level constraint on the turnover scale and a \(\sim90\%\) probability of detecting the turnover below a redshift of 0.5. Lastly, we also demonstrate that combining the DESI BGS redshift sample with the LRG, ELG, and QSO samples could break the degeneracy between \(r_H\) and \(\Omega_m\) and improve their constraints by \(\sim25\%\) and \(\sim45\%\), respectively,  compared to only using the high redshift samples. The constraints on the particle horizon at the matter-radiation equality \(r_H\) and the matter density \(\Omega_m\) could then further improve by \(\sim20\%\) and \(\sim30\%\), respectively, when combining the full set of DESI redshift tracers with 4HS and SKA1-B2.}

\begin{document}
\maketitle
\flushbottom

\section{Introduction}
The large-scale structure of the Universe provides valuable information about its contents and evolution. Since large-scale structures are seeded by primordial, approximately Gaussian density fluctuations in the early universe, one of the most popular methods for extracting information from large-scale structures is by measuring the two-point correlation function (or its Fourier counterpart, the power spectrum) of the galaxy distribution. Of interest to cosmology, the galaxy power spectrum contains information about the Baryon Acoustic Oscillations (BAO) of the early universe \citep{Eisenstein_2005, Cole_2005, Beutler_2011, Kazin_2014, Alam_2017, du_Mas_des_Bourboux_2017, Ruggeri_2020, Bautista_2020, Gil_Mar_n_2020, Raichoor_2020, de_Mattia_2020, Hou_2020, Neveux_2020, desi_key3, desi_key4}, the Redshift Space Distortions (RSD) created by the peculiar motions of galaxies \citep{Kaiser_1987, Watkins_2009, Johnson_2014, Carrick_2015, Huterer_2017, Howlett_2017, Adams_2017, Howlett_2019, Qin_2019, Adams_2020, Said_2020, Lai_2022}, and the turnover scale \citep{Blake_2005, Poole_2013, Cunnington_2022, Bahr_Kalus_2023, Bahr_Kalus_2025, Alonso_2025}. The latter is the focus of this paper. The turnover scale is related to the size of the Universe at matter-radiation equality \citep{Prada_2011}, which serves as a standard ruler. Combining this with low-redshift measurements of matter density can provide unique insight into the Hubble parameter \(H_0\). Recent observations at low redshifts using Cepheid-calibrated Type-Ia supernovae produce much higher \(H_0\) \citep{Riess_2022} than is inferred from the CMB temperature and polarisation anisotropies \citep{Tristram_2024, Planck_2020}. This discrepancy is known as the Hubble tension and remains unresolved. Since the CMB inference is tied strongly to the size of the sound horizon at the epoch of recombination, one of the most popular solutions to the Hubble tension is the early dark energy model, which alters the size of the sound horizon \citep{Sch_neberg_2022}. Combining the constraint on the turnover scale with matter density measurements at low redshift provides an alternative way to measure the Hubble parameter, independent of the sound horizon or Cepheids. Therefore, it can potentially shed some light on the Hubble tension \citep{Bahr_Kalus_2023}.

The turnover scale of the power spectrum is closely related to the time at which matter-radiation equality occurs. During the radiation-dominated era, the growth of small-scale matter perturbations that entered the horizon first was suppressed by radiation pressure. However, matter perturbations that entered the horizon after matter-radiation equality no longer experienced the radiation pressure and grew linearly. This created a distinct turnover feature in the matter power spectrum. Ref.~\citep{Poole_2013} first attempted to use the galaxy power spectrum from the WiggleZ dark energy survey to detect this turnover. However, a more concrete detection was achieved by ref.~\citep{Bahr_Kalus_2023}. Ref.~\citep{Bahr_Kalus_2025} provide constraints on the turnover scale from the Dark Energy Spectroscopic Instrument (DESI) surveys. They also combine these constraints with the DESI BAO measurements and Type Ia supernovae from Pantheon\(+\) to obtain constraints on the Hubble parameters independent of the sound horizon.    

Most previous research has focused on constraining the turnover scale at high redshifts (\(z > 0.5\)). This is mainly because cosmic variance dominates the uncertainty on the turnover scale, and is larger at low redshift due to the small cosmic volume compared to high-redshift measurements \citep{Cunnington_2022}. One method to reduce the cosmic variance limit is by combining multiple different tracers of the same underlying density field \citep{McDonald_2009}. The galaxy power spectrum is not the only tracer we can use to measure the turnover scale. Ref.~\citep{Cunnington_2022} use the HI power spectrum to forecast the constraint on the turnover scale with the MeerKLASS \citep{Santos_2017} and the Square Kilometre Array (SKA) \citep{SKA_2020} surveys. Another tracer that can potentially constrain the turnover scale is the direct measurement of galaxy peculiar velocities, which are generated through gravitational collapse. Based on linear theory, the velocity auto-power spectrum and the galaxy-velocity cross-power spectrum are proportional to the galaxy power spectrum \citep{Strauss_1995}. To measure peculiar velocity, we can use the Fundamental Plane \citep{Dressler_1987, Djorgovski_1987}, the Tully-Fisher relation \citep{Tully_1977}, and Type Ia supernovae \citep{Phillips_1993} to measure the redshift-independent distances. The difference between the redshift-independent distance measurements from these methods and the redshift measurements from spectroscopy gives the peculiar velocity. The velocity auto-power spectrum also has the added benefit of being sensitive to the larger-scale modes than the galaxy density spectrum due to the $(1/k)^2$ dependence of the velocity divergence field relative to the overdensity field. This makes them a promising probe of the turnover scale and other large-scale correlations \citep{Giani_2023}. On the other hand, measurements of the velocity power spectrum using distance indicators are limited in their redshift range. Therefore, the volume of the peculiar velocity survey will be smaller than the galaxy survey. 

In this work, we investigate whether incorporating the HI power spectrum, velocity power spectrum, and their cross-correlation with galaxy density allows us to constrain the turnover scale below a redshift of 0.5. We will constrain the turnover scale with the forecast number density of galaxy redshifts and peculiar velocities for the complete DESI BGS (Bright Galaxy Survey) and PV survey \citep{Levi_2013, Ruiz_Macias_2020, DESI_EDR, Hahn_2023, Saulder_2023} and 4HS (4-metre multi-object spectrograph telescope Hemispheric Survey) \citep{de_Jong_2012, de_Jone_2016}. We will also utilise the forecast for the SKA Medium-Deep Band 2 Survey (hereafter referred to as SKA1-B2, with its phase 2 denoted as SKA2-B2) \citep{SKA_2020} to constrain the turnover scale using the HI power spectrum. Although SKA Band 1 will cover a larger cosmic volume and provide a tighter constraint on the turnover scale \citep{Cunnington_2022}, we chose the Band 2 survey because we want to constrain the turnover scale below the redshift of 0.5. Additionally, such a measurement in the dark energy-dominated era would be interesting cosmologically, but also because one would expect such measurements to be most limited in the case of a single tracer due to the effects of sample variance. Hence, it is in this redshift range where multi-tracer approaches, which reduce the effect of sample variance as shown in ref.~\citep{McDonald_2009} are most interesting to study. We followed ref.~\citep{Bahr_Kalus_2023} to employ a model-independent approach, providing a conservative prediction for the constraints on the turnover scale of future surveys. 
\begin{comment}
    Nonetheless, in this work, we examine the effect of adding the velocity auto-power spectrum and the galaxy-velocity cross-power spectrum when constraining the turnover scale with the DESI BGS (Bright Galaxy Survey) \citep{Levi_2013, Ruiz_Macias_2020, DESI_EDR, Hahn_2023}, the 4HS (4-metre multi-object spectrograph telescope Hemispheric Survey) \citep{de_Jong_2012, de_Jone_2016}, and with LSST (Legacy Survey of Space and Time) \citep{LSST_2009}, finding that a benefit can still be realised, and that such a combined measurement of the turnover is made possible in a regime where use of only the galaxy density power spectrum would not be constraining. We followed ref.~\citep{Bahr_Kalus_2023} to employ a model-independent approach to provide a conservative prediction for the constraints on the turnover scale of future surveys. 
\end{comment}

This paper is organised as follows. In section \ref{sec:Data}, we present our method of generating mock datasets for DESI BGS, 4HS, and SKA, including the analytic covariance matrices. In section \ref{sec:Theory}, we show our method of fitting the turnover scale. In section \ref{sec:Results}, we present the constraints of the turnover scale from the simulated datasets and compare them to the constraints with only the galaxy power spectrum. We also investigate the impact of the constraints on cosmological parameters when combining our low-redshift turnover scale constraints with high-redshift ones. Lastly, we present our conclusions in section \ref{sec:Conclusion}. 

\section{Mock dataset}
\label{sec:Data}
This section explains how we generate the mock data power spectrum for each survey, which is used to obtain constraints on the turnover scale. In this work, we generate mock power spectra directly from linear perturbation theory, supplemented with additional phenomenological non-linear damping terms, rather than using N-body simulations. This is appropriate because the constraint on the turnover mainly depends on the shape of the power spectrum on large scales, which linear perturbation theory can model accurately, and allows us to switch between different survey number densities without having to regenerate a large number of simulations. The fiducial cosmological parameters for the eBOSS mock dataset are dark matter density \(\Omega_{\mathrm{cdm}}h^2 = 0.1197\), baryon density \(\Omega_bh^2 = 0.022\), primordial amplitude of the power spectrum \(A_s = 2.0341\times10^{-9}\), and the Hubble parameter at redshift zero \(H_0 = 67.60 \mathrm{km} \mathrm{s}^{-1}\mathrm{Mpc}^{-1}\) to match the fiducial cosmology of ref.~\citep{Bahr_Kalus_2023}. For other mock datasets, we use the baseline \(\Lambda\)CDM cosmology from \textsc{AbacusSummit} \citep{Garrison_2021, Maksimova_2021} with \(\Omega_{\mathrm{cdm}}h^2 = 0.1200\), \(\Omega_bh^2 = 0.02237\), \(A_s = 2.0830\times10^{-9}\), and \(H_0 = 67.36 \mathrm{km} \mathrm{s}^{-1}\mathrm{Mpc}^{-1}\).   

When combining galaxy redshifts and peculiar velocities in our analysis, we adopted the quasi-linear model from ref.~\citep{Koda_2014}, where the galaxy auto-power spectrum, velocity auto-power spectrum, and the galaxy-velocity cross-power spectrum are given by 
\begin{align}
    P_{gg}(k, \mu, \sigma_g) &= (b+f\mu^2)^2 D_g^2(k, \mu, \sigma_g) P_{mm}(k),
    \label{eq: P_gg_mu} \\
    P_{vv}(k, \mu, \sigma_u) &= \left(\frac{aHf\mu}{k}\right)^2 D_u(k, \sigma_u)^2 P_{\theta \theta}(k),
    \label{eq: P_vv_mu} \\
    P_{gv}(k, \mu, \sigma_g, \sigma_u) &= \frac{iaHf\mu}{k}(b + f\mu^2) D_g(k, \mu, \sigma_g) D_u(k, \sigma_u) P_{m \theta}(k).
    %\label{eq:P_gv_mu}
\end{align}
Here, \(D_g(k, \mu, \sigma_g)\) models the small-scale finger-of-god effect on the density field
\begin{equation}
    D_g = e^{-\frac{(k \mu \sigma_g)^2}{2}},
    \label{eq:D_g}
\end{equation}
where \(\sigma_g\) is the damping factor that we can tune to represent a particular set of galaxies and \(\mu = \hat{\boldsymbol{k}} \cdot \hat{\boldsymbol{r}}\) is the cosine of the angle between wavevector \(\hat{\boldsymbol{k}}\) and the line of sight direction \(\hat{\boldsymbol{r}}\).\footnote{Throughout this work, we use bold letters to indicate vector quantities.} Similarly, the modelling of the finger-of-god damping on the velocity field is given by \citep{Koda_2014, Adams_2020}
\begin{equation}
    D_u = \frac{\sin{(k\sigma_u)}}{k\sigma_u}, 
    \label{eq: D_u}
\end{equation}
where \(\sigma_u\) is the damping factor specifically for the velocity field. Additionally, \(P_{mm}\), \(P_{m\theta}\), and \(P_{\theta \theta}\) are the matter power spectrum, the matter-velocity divergence cross-power spectrum, and the velocity divergence auto-power spectrum. Lastly, \(H\) is the Hubble parameter, \(a\) is the scale factor, \(f\) is the logarithmic growth rate, and \(b\) is the linear galaxy bias given by 
\begin{equation}
    b(z) = b_0/D(z).
    \label{eq:b_z}
\end{equation}
\(D(z)\) is the linear growth factor normalized to \(D(z=0) = 1\). The linear bias can be chosen given a fiducial cosmological model and to represent a particular class of galaxies. When compared to the power spectrum from simulations, this model is shown to be accurate up to \(k_{\mathrm{max}} = 0.20 h \mathrm{Mpc}^{-1}\) \citep{Koda_2014}.

The HI power spectrum is obtained by considering the two-point clustering statistics in the three-dimensional Fourier space measurements obtained through HI Intensity Mapping. We adopted the models from ref.~\citep{Bull_2015} and \citep{Masui_2013, Villaescusa-Navarro_2015, Blake_2019} for the auto and cross-power spectra between temperature and galaxy density 
\begin{align}
    P_{TT} &= (b_{HI} + f\mu^2)^2D_T^2P_{mm}(k).
    \label{eq:P_TT} \\
    P_{gT} & = (b_{HI} + f\mu^2)(b + f\mu^2)D_TD_gP_{mm}(k).
    \label{eq:P_gT}
\end{align}
The HI auto-power spectrum and the HI-galaxy cross-power spectrum are given by 
\begin{align}
    P_{HI} &= \bar{T}_{HI}^2P_{TT}.
    \label{eq:P_HI} \\
    P_{gHI} & = \bar{T}_{HI}T_{gT}.
    \label{eq:P_gHI}
\end{align}
The damping factor \(D_T\) is 
\begin{equation}
    D_T = e^{-\frac{(k \mu \sigma_{NL})^2}{2}},
    \label{eq:D_T}
\end{equation}
similar to equation~(\ref{eq:D_g}) but with a different damping factor \(\sigma_{NL}\). We follow ref.~\citep{Li_2007, Bull_2015} to set \(\sigma_{NL} = 7 h^{-1} \mathrm{Mpc}\). 

We make use of the best-fit power law models in ref.~\citep{Santos_2015} for the redshift evolution of the mean HI temperature and the HI bias 
\begin{align}
    \bar{T}_{HI}(z) &= 0.055919 + 0.23242z - 0.024136z^2
    \label{eq:T_HI} \\
    b_{HI}(z) &= (b_{HI}^0 / 0.677105)(0.66655 + 0.17765z + 0.050223z^2),
    \label{eq:b_HI}
\end{align} 
where \(b_{HI}^0\) is the HI bias at redshift zero.\footnote{These equations are taken from the \textsc{bao21cm} code \citep{Bull_2015}. See \url{https://gitlab.com/radio-fisher/bao21cm/-/tree/master?ref_type=heads} for more details.} Although there are different parameterisations for the HI bias and mean temperature in the literature, we show that these different parameterisations have no impact on the constraints on the turnover scale in Appendix \ref{sec:HI_model}. 

We generate \(P_{mm}\), \(P_{\theta \theta}\), and \(P_{vv}\) from the code \textsc{Copter}\footnote{\url{https://github.com/jwgcarlson/Copter}} \citep{Carlson_2009} using two-loop Renormalized Perturbation Theory \citep{Crocce_2006a, Crocce_2006b, Crocce_2008}. We use \textsc{CAMB} \citep{Lewis_2000} to calculate the growth rate \(f\) at the effective redshift of the survey, and the galaxy bias at redshift zero \(b_0\) is taken from Table.~\ref{tab:survey_param}. For eBOSS, the chosen bias parameter is consistent with constraints from ref.~\citep{Hou_2020, Neveux_2020, Chudaykin_2023} at the effective redshift. The DESI bias parameter is consistent with ref.~\citep{DESI_2016a} at redshift zero. For the other two surveys, the bias parameter is consistent with ref.~\citep{Whitford_2022}. To generate the mock data, we take \(\sigma_g = 5.8h^{-1}\mathrm{Mpc}\) and \(\sigma_u = 13.0h^{-1}\mathrm{Mpc}\) from ref. \citep{Koda_2014}. Lastly, we generate the data power spectrum at the effective redshift of the survey. Based on equation~(\ref{eq:cov_full}) and the results given in more detail in the next section, the effective volume of the survey for tracer \(a\) is given by 
\begin{align}
    V_{\mathrm{eff}}^a(k) &= \Omega_{\mathrm{sky}}^a \int dz D_c^2 \frac{d D_c}{dz}\left(\frac{P_{aa}(k, z)}{P_{aa}(k,z) + \frac{\Sigma_a(z)}{\Bar{n}_a(z)}}\right)^2 \nonumber \\
    &= \Omega_{\mathrm{sky}}^a \int dz V_{\mathrm{int}}(k, z),
    \label{eq: V_eff}
\end{align}
where the noise term \(\Sigma_a\) and number density \(\bar{n}_a\) are defined between equations~(\ref{eq: n_T}) and (\ref{eq:sigma_TT}).
The effective redshift \(z_{\mathrm{eff}}\) is then defined as
\begin{align}
    z_{\mathrm{eff}}^a(k) &= \frac{\Omega_{\mathrm{sky}}^a \int dz V_{\mathrm{int}}(k, z)\times z}{V_{\mathrm{eff}}^a(k)} \nonumber \\
    &= \frac{V_z^a(k)}{V_{\mathrm{eff}}^a(k)}.
    \label{eq:z_eff}
\end{align}
In this work, instead of taking the volume average in Fourier space of \(z_{\mathrm{eff}}(k)\), we use \(z_{\mathrm{eff}}(k_{\mathrm{TO}}^{\mathrm{fid}})\) as the effective redshift, where \(k_{\mathrm{TO}}^{\mathrm{fid}}\) is the fiducial turnover scale based on the fiducial cosmological parameters. This is because we want to maximise the signal-to-noise around the turnover scale to obtain the tightest constraints. Similarly, for a multi-tracer analysis, we consider the effective redshift as the redshift that maximises the signal-to-noise of the measurement, the effective redshift when combining two tracers can approximately be given by
\begin{equation}
    z_{\mathrm{eff}}^{a, b}(k) = \frac{V_z^a(k) + V_z^b(k)}{V_{\mathrm{eff}}^a(k) + V_{\mathrm{eff}}^b(k)}, 
    \label{eq:z_eff_2}
\end{equation}
where we have neglected the contribution of the cross-power spectrum in defining our effective volumes.

Table.~\ref{tab:survey_param} and \ref{tab:survey_SKA_param} summarise the survey properties we used to generate the mock power spectrum. The same parameters were used to estimate the Gaussian analytical covariance matrix for these surveys in section \ref{sec:covariance}, providing semi-realistic uncertainties for our mock measurements. Although ref.~\citep{Braun_2015, Bonaldi_2016} provide the expected total integration time and survey area for SKA2-B2, they do not provide a forecast for the system noise temperature and number of dishes for SKA2-B2. Therefore, we consider a pessimistic scenario here where SKA2-B2 has the same system noise temperature and the number of dishes as SKA1-B2. For comparison with real measurements of the turnover scale from the eBOSS QSO survey, we also utilise the number density files and survey properties from ref.~\citep{Hou_2020, Bahr_Kalus_2023}. We will use these to demonstrate that the constraints on the turnover scale from our mock power spectra and covariance matrix are consistent with those of the real data \citep{Bahr_Kalus_2023}. Lastly, we include the DESI LRG, ELG, and QSO samples to illustrate how low redshift constraints on the turnover scale complement those from high redshift in section \ref{sec:combined}.  

\begin{figure}
    \centering
    \includegraphics[width=0.70\textwidth]{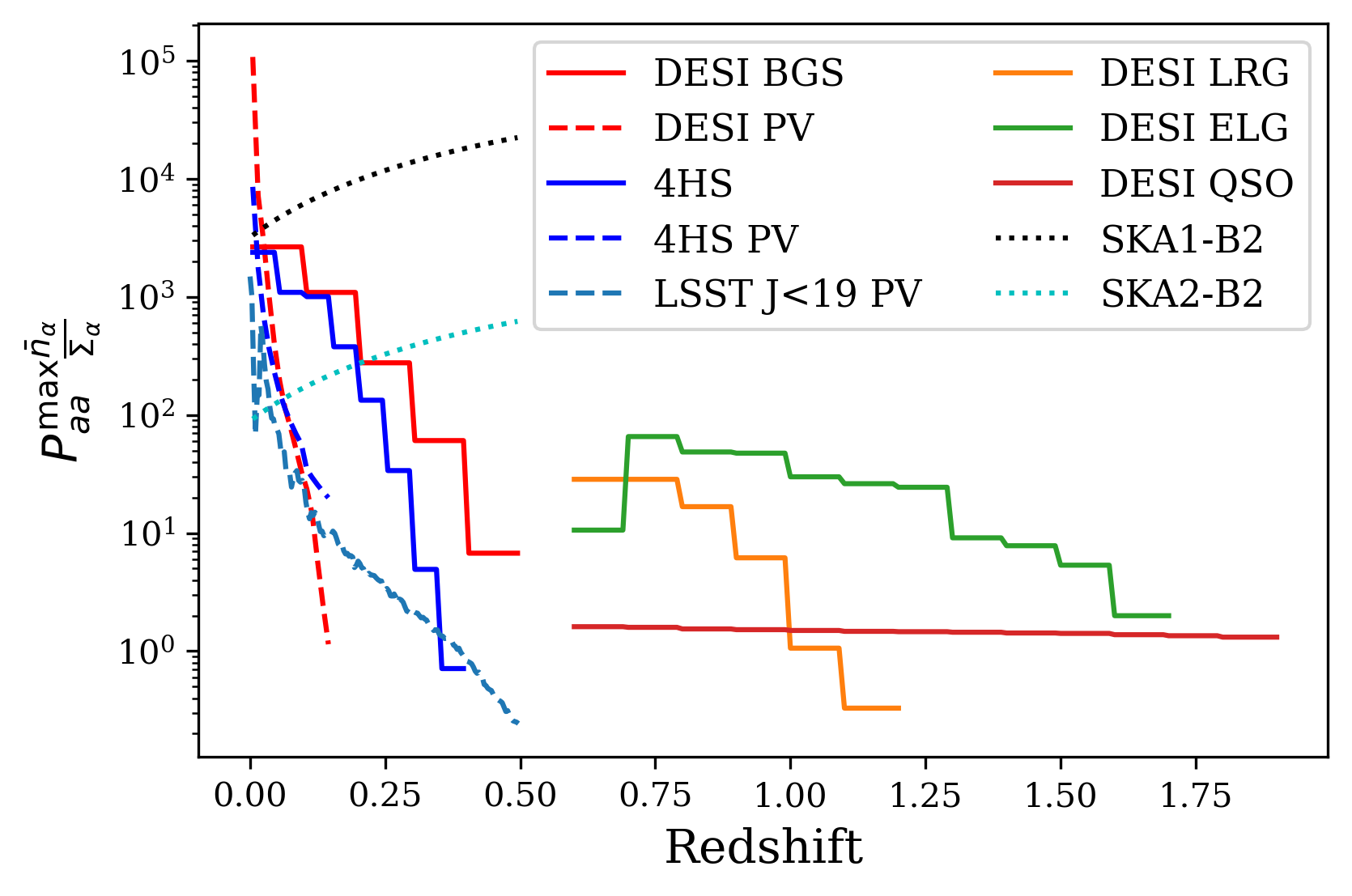}
    
    \caption{\(P_{aa}^{\mathrm{max}} \frac{\bar{n}_a}{\Sigma_a}\) for all surveys in Table.~\ref{tab:survey_param} and \ref{tab:survey_SKA_param}. The solid lines are galaxy surveys, dashed lines are peculiar velocity surveys, and dotted lines are intensity mapping surveys. For different tracers within the same survey, they also have the same colour. For the cosmic signal to dominate over the noise, we want \(P_{aa}^{\mathrm{max}} \frac{\bar{n}_a}{\Sigma_a} \gg 1\), which is true for most surveys. However, we caution the reader that a survey with a higher \(P_{aa}^{\mathrm{max}} \frac{\bar{n}_a}{\Sigma_a}\) does not mean it will have tighter turnover scale constraints than the survey with a lower value since the parametrisation here does not fully account for the impact of survey volume on the parameter constraints.}
    \label{fig:nbar_plot}
\end{figure}

Based on equation~(\ref{eq: V_eff}), we want \(P_{aa}(k, z) \frac{\bar{n}_a(z)}{\Sigma_a(z)} \gg 1\) such that the cosmic signal dominates the noise. Fig.~\ref{fig:nbar_plot} demonstrates that cosmic signals from most surveys in Table.~\ref{tab:survey_param} and \ref{tab:survey_SKA_param} dominate over the noise. \(P_{aa}^{\mathrm{max}}\) denotes the maximum amplitude of the power spectrum at the effective redshift. This is used because the turnover scale constraint mainly depends on the signal-to-noise of the power spectrum around its maximum amplitude. However, we caution the reader that a survey with a higher \(P_{aa}^{\mathrm{max}} \frac{\bar{n}_a}{\Sigma_a}\) does not mean it will have tighter turnover scale constraints than the survey with a lower value. For example, Fig.~\ref{fig:nbar_plot} illustrates that SKA1-B2 has a higher \(P_{aa}^{\mathrm{max}} \frac{\bar{n}_a}{\Sigma_a}\) than SKA2-B2 despite section~\ref{sec:Results} showing SKA2-B2 will provide stronger constraints than SKA1-B2. This is because \(P_{aa}^{\mathrm{max}} \frac{\bar{n}_a}{\Sigma_a}\) represents the signal-to-noise per mode and does not take the impact of survey volume on subsequent constraints into account. Our SKA2-B2 configuration in Fig.~\ref{fig:nbar_plot} has a lower signal-to-noise per mode than SKA1-B2 because we are (pessimistically) considering a phase-2 survey with the same integration time, number of dishes, and system temperature as phase-1, but covering a larger area on the sky. As can be seen in equations~(\ref{eq: n_T}) and~(\ref{eq:sigma_TT}), this results in a lower effective ``number density'' of tracers and a larger temperature noise per unit sky area. 

\begin{comment}
\begin{figure}
    \centering
    \includegraphics[width=0.495\textwidth]{Pictures/Galaxy_nbar.png}
    \includegraphics[width=0.495\textwidth]{Pictures/Velocity_nbar.png}
    
    \caption{Left: the average number density of galaxies for DESI, 4HS, and LSST. The maximum redshifts for DESI and LSST are 0.5, and the maximum redshift for 4HS is 0.4. Right: The average number density for peculiar velocities for DESI, 4HS, and LSST.}
    \label{fig:nbar_plot}
\end{figure}
\end{comment}
\begin{comment}
Fig.~\ref{fig:nbar_plot} illustrates the number density of galaxies we assume for DESI, 4HS, and LSST on the left and the number density of peculiar velocities for the same surveys on the right. DESI and 4HS have a higher number density of peculiar velocities than LSST because supernovae are rarer than elliptical and spiral galaxies. In contrast, LSST can measure peculiar velocities to higher redshifts because supernovae provide more accurate measurements of peculiar velocities (the relative uncertainty we assume for the PVs from each survey are given in Table~\ref{tab:survey_param}, and enters into our covariance matrix estimation in section \ref{sec:covariance}). We use the same number density as ref. \citep{Whitford_2022}, so we refer the readers to section 4 of that paper for more information on these surveys. 
\end{comment}

\begin{table}[t!]
% title of Table
     % is used to refer this table in the text
\centering                          % used for centering table
\renewcommand{\arraystretch}{1.4} % Default value: 1
\begin{tabular}{c|c|c|c|c}        % centered columns (4 columns)
%\hline\hline                 % inserts double horizontal lines    % table heading 
Survey & $\Omega_{\mathrm{sky}}$ ($\mathrm{deg}^2$) & $b_0$ &  PV relative uncertainty & Reference\\ \hline \hline
eBOSS &  4699 & 1.68 & N.A & \citep{Hou_2020, Bahr_Kalus_2023} \\
\hline
DESI BGS + PV & 14000 & $1.34$ & 0.2 & \citep{DESI_Valid,Saulder_2023}\\
\hline
DESI LRG & 14000 & $1.70$ & N.A & \citep{DESI_Valid, DESI_2016a} \\
\hline
DESI ELG & 14000 & $0.84$ & N.A & \citep{DESI_Valid, DESI_2016a} \\ 
\hline
DESI QSO & 14000 & $1.20$ & N.A & \citep{DESI_Valid, DESI_2016a} \\
\hline
4HS + PV & 17000 & $1.34$ & 0.2 & \citep{Whitford_2022}\\
\hline
LSST J < 19 sample & 18000 & N.A &0.05 &\citep{Howlett_2017b, Whitford_2022}
\end{tabular}
\caption{The survey parameters used to generate the covariance matrix for galaxy density and peculiar velocity. The LSST J < 19 sample is a predicted number density of peculiar velocities measured based on the number of type Ia supernovae that are expected to be detected by LSST and that could also have host galaxy
redshifts \citep{Howlett_2019, Whitford_2022}.}    
\label{tab:survey_param} 
\end{table}

\begin{table}[t!]
% title of Table
     % is used to refer this table in the text
\centering                          % used for centering table
\renewcommand{\arraystretch}{1.4} % Default value: 1
\begin{tabular}{c|c|c|c|c|c|c|c}        % centered columns (4 columns)
%\hline\hline                 % inserts double horizontal lines    % table heading 
Survey & $\Omega_{\mathrm{sky}}$ ($\mathrm{deg}^2$) & $b_{HI}^0$ & $T_{\mathrm{rx}}$ ($K$) & $N_{\mathrm{dish}}$ & $t_{\mathrm{tot}}$ (Hours) & $\Delta z$ &Reference\\ \hline \hline
SKA1-B2 &  5000 & 0.677 & 7.50 & 197 &10000 & 0 - 0.5 & \citep{SKA_2020} \\
\hline
SKA2-B2 & 30000 & 0.677 & 7.50 & 197 &10000 & 0 - 0.5 & \citep{Braun_2015, Bonaldi_2016} \\
\end{tabular}
\caption{The survey parameters used to generate the covariance matrix for the HI power spectrum. SKA1-B2 denotes the SKA phase one (medium-deep) band two survey \citep{SKA_2020}, and SKA2-B2 denotes the proposed SKA phase two band two survey \citep{Braun_2015, Bonaldi_2016}, assuming it has similar systematics as phase 1. The HI bias at redshift zero is taken from ref.~\citep{Bull_2015}. \(\Delta z\) here denotes the redshift range of the survey.}    
\label{tab:survey_SKA_param} 
\end{table}

\subsection{Analytical covariance matrix}
\label{sec:covariance}
This section describes how we produce semi-realistic uncertainties for our mock datasets, under the Gaussian limit (i.e., neglecting higher-order clustering terms that enter the full covariance matrix) and ignoring the more complex effects of the survey window function (we assume a uniform sky completeness over the survey area, and a selection function that only varies with redshift/distance). We will refer the readers to the Appendix \ref{sec:anal_cov} for the derivation of the analytical covariance matrix and only show the final expression here. In general, the Gaussian covariance matrix between two power spectrum multipoles $P_{\ell_{1}}^{ab}(k)$ and $P_{\ell_{2}}^{cd}(k^{\prime})$ is given by 
\begin{multline}
    C^{abcd}_{l_1l_2}(k_1,k_2) = \frac{(2\pi)^3(2l_1+1)(2l_2+1)}{2V_kI_{ab}I_{cd}}\delta^{D}(k_1-k_2) \Omega_{\mathrm{tracer}}\int_{-1}^{1}d\mu \mathcal{L}_{l_1}(\mu) \mathcal{L}_{l_2}(\mu) \int dz\, D_{c}^{2}(z)\frac{dD_{c}}{dz} \\ w_a(z) w_b(z) w_c(z) w_d(z) \biggl[\mathcal{P}_{ad}(k_1,\mu,z)\mathcal{P}_{bc}(k_1,\mu,z) + \mathcal{P}_{ac}(k_1,\mu,z)\mathcal{P}_{bd}(k_1,\mu,z)\biggl], 
    \label{eq:cov_full}
\end{multline}
where \(a, b, c, d\) could either be the galaxy density \(g\),  velocity \(v\), or temperature \(T\) fields.
The normalisation of the individual power spectra is represented by 
\begin{equation}
    I_{\alpha \beta} = \Omega_{\mathrm{tracer}}\int dz\,D_{c}^{2}(z)\frac{dD_{c}}{dz}  n_{\alpha}(z)n_{\beta}(z) w_{\alpha}(z)w_{\beta}(z),
    \label{eq: G_xy_1}
\end{equation}
where $D_{c}(z)$ is the comoving radial distance to redshift $z$ and \(\mathcal{L}_l\) is the Legendre polynomial with order \(l\). The kernel \(\mathcal{P}\) is given by 
\begin{equation}
    \mathcal{P}_{\alpha \beta}(k, \mu, z) = P_{\alpha\beta}(k, \mu, z) \bar{n}_\alpha(z) \bar{n}_\beta(z) + \bar{n}_\alpha(z) \Sigma_\alpha(z) \delta_{\alpha}^{\beta},
\end{equation}
where \(\delta_{\alpha}^{\beta}\) is the kronecker delta function.
\begin{comment}
If the superscripts for \(\mathcal{P}\) are the same, then it is given by
\begin{equation}
    \mathcal{P}^{\alpha \alpha} = \Bar{n}_\alpha^2 P_{\alpha\alpha} + \Bar{n}_\alpha \Sigma_\alpha.
    \label{eq: P_auto}
\end{equation}
If the superscripts for \(\mathcal{P}\) are different, then it is given by 
\begin{equation}
    \mathcal{P}^{\alpha\beta} = P_{\alpha\beta} \Bar{n}_\alpha \Bar{n}_\beta. 
    \label{eq:P_cross}
\end{equation}
\end{comment}
The expression for the auto-power spectrum \(P_{\alpha\alpha}\) and the cross-power spectrum \(P_{\alpha\beta}\) are given between equation~(\ref{eq: P_gg_mu}) and equation~(\ref{eq:P_gT}) for different tracers. The number density \(\Bar{n}_\alpha\) and the noise term \(\Sigma_\alpha\) are different for different tracers, and their derivations are shown in Appendix \ref{sec:anal_cov}. For the galaxy density field and the peculiar velocity field, it will be just the number density \(\Bar{n}_g\) and \(\Bar{n}_v\), respectively, of measured galaxy redshifts and peculiar velocities. For the temperature field, its effective ``number density'' is given by 
\begin{equation}
    \bar{n}_T(z) = \frac{\bar{T}_{HI}(z)}{V_{HI}},
    \label{eq: n_T}
\end{equation}
where \(V_{HI}\) is the volume of the HI survey \citep{Blake_2019}. For the noise term, 
\begin{align}
    \Sigma_g(z) &= 1
    \label{eq: sigma_gg} \\
    \Sigma_v(z) &= \sigma_{v, \mathrm{err}}^2(z) + \sigma_{v, \mathrm{nl}}^2
    \label{eq:sigma_VV} \\
    \Sigma_T(z) &= \frac{\varsigma_T^2(z)}{\bar{T}_{HI}(z)},
    \label{eq:sigma_TT}
\end{align}
for the galaxy density field, the peculiar velocity field, and the temperature field, respectively. These represent additional sources of noise entering the power spectrum from, e.g., intrinsic scatter in the empirical relationships used to estimate peculiar velocities \citep{Howlett_2019}, or noise in he temperature field arising from instrumental effects \citep{Bull_2015}. We show the detailed derivation of the noise terms in Appendix \ref{sec:anal_cov}. We adopt \citep{Howlett_2019} 
\begin{equation}
    \sigma_{v, \mathrm{err}}(z) = \Delta_v H_0D_c(z)
    \label{eq:sigma_v_err}
\end{equation}
for the uncertainty in the peculiar velocity measurements, where \(\Delta_v\) denotes the relative peculiar velocity uncertainty in Table.~\ref{tab:survey_param} and 
\begin{equation}
    \sigma_{v, \mathrm{nl}} = 300 \mathrm{km/s}
    \label{eq:sigma_v_nl}
\end{equation}
denotes the nonlinear velocity dispersion. Lastly, we adopt \citep{Bull_2015} 
\begin{equation}
    \varsigma_T(z) = \frac{4T_{\mathrm{sys}}(z)}{\eta\pi}\sqrt{\frac{\Omega_{\mathrm{sky}}}{n_{\mathrm{pol}}t_{\mathrm{tot}}\nu_{21}N_{\mathrm{d}}N_{\mathrm{b}}}}. 
\end{equation}
Here, the number of beams per dish \(N_{\mathrm{b}} = 1\) \citep{Bull_2015}, the rest frame frequency of the 21cm line is \(\nu_{21} = 1420.406 \mathrm{MHz}\), the number of supported polarization channel is \(n_{\mathrm{pol}} = 2\), the survey efficiency \(\eta = 0.7\) \citep{Bull_2015}, and the system temperature is given by 
\begin{equation}
    T_{\mathrm{sys}} = T_{\mathrm{rx}} + T_{\mathrm{spl}} + T_{\mathrm{CMB}} + T_{\mathrm{gal}},
    \label{eq: T_sys}
\end{equation}
where the spill-over contribution is \(T_{\mathrm{spl}} = 3K\) \citep{SKA_2020}, the CMB temperature \(T_{\mathrm{CMB}} = 2.73K\), and the noise contribution from our own galaxy is modelled by \citep{SKA_2020}
\begin{equation}
    T_{\mathrm{gal}} = 25K (408\mathrm{MHz}(1 + z)/\nu_{21})^{2.75}.
    \label{eq: T_gal}
\end{equation}
The receiver noise temperature \(T_{\mathrm{rx}}\), the number of dishes \(N_{\mathrm{d}}\), and the total integration time \(t_{\mathrm{tot}}\) are given in Table.~\ref{tab:survey_SKA_param}. The model for \(\varsigma_T\) here neglects the effect of instrumental beams and the residual foreground \citep{Bull_2015}. In Appendix \ref{sec:HI_noise}, we show that both effects have little impact on the turnover scale constraints because the instrumental beam mainly affects the small-scale power spectrum \citep{Blake_2019} and the residual foreground is expected to be small \citep{Bull_2015}. 

Following our definitions of the various number densities and additional noise terms, the optimal weight for any tracer \(\alpha\) is given by
\begin{equation}
    w_{\alpha}(k, \mu, z) = \frac{1}{\bar{n}_\alpha(z) \left(P_{\alpha\alpha}(k, \mu, z) + \frac{\Sigma_\alpha(z)}{\Bar{n}_\alpha(z)}\right)}. 
    \label{eq:FKP}
\end{equation}
Substituting equation~(\ref{eq: sigma_gg}) into equation~(\ref{eq:FKP}), we will recover the well-known FKP weight for the galaxy redshift surveys \citep{Feldman_1994}. Furthermore, if we substitute equation~(\ref{eq:sigma_VV}) and equations~(\ref{eq:sigma_TT}) and (\ref{eq: n_T}) into equation~(\ref{eq:FKP}), we recover the weight for the velocity power spectrum \citep{Howlett_2019} and the HI intensity mapping \citep{Blake_2019}, respectively. Usually, a constant power spectrum amplitude \(P^{\alpha\alpha}_{\mathrm{FKP}}\) is used in equation~(\ref{eq:FKP}) instead of the power spectrum. We follow ref.~\citep{Qin_2019} to set \(P^{gg}_{\mathrm{FKP}} = 1600 h^{-3} \mathrm{Mpc}^3\) and \(P^{vv}_{\mathrm{FKP}} = 5\times10^9 h^{-3}\mathrm{Mpc}^3\mathrm{km}^{2}\mathrm{s}^{-2}\). The weight for the HI tracer can be rewritten as \citep{Blake_2019}
\begin{align}
    w_{T}(k, \mu, z) &= \frac{\bar{T}_{HI}(z)}{\frac{P_{TT}(k, \mu, z)}{V}\bar{T}_{HI}^2(z) + \varsigma_T^2(z)} \nonumber \\
    &= \frac{\bar{T}_{HI}(z)}{\frac{P_{HI}(k, \mu, z)}{V} + \varsigma_T^2(z)},
    \label{eq:w_HI} 
\end{align}
and we set \(P^{HI}_{FKP} = 300 h^{-3} \mathrm{Mpc}^3 mK^2\), which is close to the value of our HI power spectrum at the turnover scale at the effective redshift. We found that our constraints on the turnover scale with the true optimal weight are the same as those with the constant weight. 

For the auto-covariance of the auto-power spectrum (\(a=b=c=d\)) or the cross-covariance of the auto-power spectrum (\(a=b\) and \(c=d\)), \(\Omega_{\mathrm{tracer}} = \Omega_{\mathrm{sky}}\) in Table.~\ref{tab:survey_param} and \ref{tab:survey_SKA_param}. Otherwise, \(\Omega_{\mathrm{tracer}} = \Omega_{\mathrm{overlap}}\), where \(\Omega_{\mathrm{overlap}}\) is the overlapping area between the two tracers. For simplicity, we categorise all surveys into two groups. Northern sky surveys and Southern sky surveys. DESI will be a Northern sky survey, LSST, 4HS, and SKA1-B2 will be Southern sky surveys. For surveys within the same hemisphere, we assume the overlapping sky area will be the sky area of the smaller survey. We also assume the Northern sky surveys have no overlapping area with the Southern sky surveys. 

\subsection{Determining \(k_{\mathrm{min}}\) for different power spectra}
To detect the turnover scale of the power spectrum, we need to be able to measure separations between pairs of galaxies on a scale larger than the expected turnover scale. The maximal separation is dictated by the geometry and maximum redshift of a particular survey, which corresponds to a particular minimum $k$-scale, $k_{\mathrm{min}}$, below which we cannot hope to measure the power spectrum. In other words, a survey needs to have a large enough volume for us to measure the turnover scale, and this fact must be taken into account when creating our mock data. This consideration is complicated by the fact that in this work, we consider different tracers from different surveys, which may have different maximum redshifts. For example, the available pairs with which to measure the velocity auto-power spectrum permit a smaller maximal separation than for the density-velocity cross-power spectrum, which itself is limited to smaller scales than the density auto-power spectrum. Therefore, we must compute and include in our analysis different values of $k_{\mathrm{min}}$ for each of these spectra, for each survey configuration.

To compute the relevant limiting scales, we follow ref. \citep{Bahr_Kalus_2023} to calculate the maximum radial distance (\(r_{\parallel}^{\mathrm{max}}\)) and the maximum angular distance (\(r_{\perp}^{\mathrm{max}}\)) for our mock data with 
\begin{equation}
    r_{\parallel}^{\mathrm{max}} = D_c(z^{\mathrm{max}}) - D_c(z^{\mathrm{min}})
    \label{eq: r_parallel}
\end{equation}
and 
\begin{equation}
    r_{\perp}^{\mathrm{max}} = \sin{\left(\frac{\sqrt{A}}{2}\right)}\left[D_c(z^{\mathrm{max}}_a) + D_c(z^{\mathrm{max}}_b)\right],
    \label{r_perp}
\end{equation}
where \(z^{\mathrm{max}} = \max{(z_a^{\mathrm{max}}, z_b^{\mathrm{max}})}\) and \(z^{\mathrm{min}} = \min{(z_a^{\mathrm{min}}, z_b^{\mathrm{min}})}\) for the cross-power spectrum. Here, \(z_a^{\mathrm{max/min}}\) is the maximum/minimum redshift for tracer \(a\), and \(A\) is the (overlapping) angular area of the survey in steradians. We then calculate the maximum volume-averaged distance with 
\begin{equation}
    D_V^{\mathrm{{max}}} = (r_{\perp}^{\mathrm{max}})^{\frac{2}{3}} (r_{\parallel}^{\mathrm{max}})^{\frac{1}{3}}. 
\end{equation}
Our definition of \(D_V^{\mathrm{{max}}}\) coincides with the definition of \(D_V\) in equation~(\ref{eq: D_v}) where we replace \(D_H\) with \(r_{\parallel}\) as the distance along the line-of-sight and \((1+z)^2D_A^2\) with \(r_{\perp}\) as the distance perpendicular to the line-of-sight. Lastly, the \(k_{\mathrm{min}}\) of the survey is approximated as 
\begin{equation}
    k_{\mathrm{min}} = \frac{2\pi}{D_V^{\mathrm{max}}}. 
    \label{eq:kmin}
\end{equation}
Once we determine the \(k_{\mathrm{min}}\), we define it as the lower boundary of the first wavenumber bin. We set the width of the bin to be \(\Delta k = 0.005 h\mathrm{Mpc}^{-1}\).\footnote{We chose this bin width since it is one of the most commonly used bin widths in galaxy surveys. We also found that our turnover scale constraints are independent of the bin width, as expected.} The number of bins is determined such that the upper boundary of the last bin is below \(0.20 h\mathrm{Mpc}^{-1}\). We compute the model and the data power spectra at the centre of each bin \(k_{\mathrm{mid}}\) with 
\begin{equation}
    P(k_{\mathrm{mid}}) = \frac{1}{V_k}\int_{k_{\mathrm{mid}}-\frac{\Delta k}{2}}^{{k_{\mathrm{mid}}+\frac{\Delta k}{2}}} dk  k^2P(k). 
\end{equation}
For multi-tracer approaches with different \(k_{\mathrm{min}}\) values, we consider two distinct cases. If the larger \(k_{\mathrm{min}}\) is still smaller than the upper boundary of the first wavenumber bin, then we use the smaller \(k_{\mathrm{min}}\) as the \(k_{\mathrm{min}}\) for both tracers.\footnote{We also tried to set the \(k_{\mathrm{min}}\) for both tracers to be the larger \(k_{\mathrm{min}}\) and found it has no impact on our constraints. Firstly, the covariance matrix is sensitive to the \(k_{\mathrm{min}}\). For a larger \(k_{\mathrm{min}}\), the covariance matrix will also be smaller, partially mitigating the effect of using a larger \(k_{\mathrm{min}}\) to constrain the turnover scale. Secondly, the first \(k_{\mathrm{mid}}\) generally has the largest relative uncertainty, so its contribution to the turnover scale constraints is small.} Otherwise, we cut all \(k\) for the tracer with the lower \(k_{\mathrm{min}}\) until \(k_{\mathrm{min}} \geq k_{\mathrm{mid}} - \frac{\Delta k}{2}\). 

\begin{table}[t!]
% title of Table
     % is used to refer this table in the text
\centering                          % used for centering table
\renewcommand{\arraystretch}{1.4} % Default value: 1
\begin{tabular}{c|c|c|c|c}        % centered columns (4 columns)
%\hline\hline                 % inserts double horizontal lines    % table heading 
Survey & $k_{\mathrm{min}}^{gg}$ & $k_{\mathrm{min}}^{TT}$ & $k_{\mathrm{min}}^{vv}$ & $k_{\mathrm{TO}}^{\mathrm{fid}}$ \\ \hline \hline
DESI BGS + PV & $ 0.0033 h \mathrm{Mpc}^{-1}$ & N.A & $ 0.0101 h \mathrm{Mpc}^{-1}$ & $0.0165 h\mathrm{Mpc}^{-1}$\\
\hline
4HS + PV & $0.0039 h\mathrm{Mpc}^{-1}$ & N.A & $0.0097h\mathrm{Mpc}^{-1}$ & $0.0165 h\mathrm{Mpc}^{-1}$\\
\hline
LSST J < 19 sample & N.A & N.A & $0.0032 h \mathrm{Mpc}^{-1}$ & $0.0165 h\mathrm{Mpc}^{-1}$\\
\hline
DESI LRG & $ 0.0023 h \mathrm{Mpc}^{-1}$ & N.A & N.A & $0.0165 h\mathrm{Mpc}^{-1}$\\
\hline
DESI ELG & $ 0.0017 h \mathrm{Mpc}^{-1}$ & N.A & N.A & $0.0165 h\mathrm{Mpc}^{-1}$\\
\hline
DESI QSO & $ 0.0016 h \mathrm{Mpc}^{-1}$ & N.A & N.A & $0.0165 h\mathrm{Mpc}^{-1}$\\
\hline
SKA1-B2 & N.A & $0.0043 h \mathrm{Mpc}^{-1}$ & N.A & $0.0165 h\mathrm{Mpc}^{-1}$ \\
\hline
SKA2-B2 & N.A & $0.0030 h \mathrm{Mpc}^{-1}$ & N.A & $0.0165 h\mathrm{Mpc}^{-1}$
\end{tabular}
\caption{The \(k_{\mathrm{min}}\) for different tracers for DESI, 4HS, LSST J < 19 sample, SKA1-B2, and SKA2-B2. For DESI BGS and 4HS, we set the maximum redshift to 0.15 for peculiar velocity \citep{Taylor_2023}. The $k_{\mathrm{min}}^{vv}$ is very close to the fiducial turnover scale, so peculiar velocity surveys alone may not be able to constrain the turnover scale. However, the cross-correlation between the galaxy and velocity could potentially probe larger scales than the velocity auto-power spectrum and provide constraints on the turnover scale.}    
\label{tab:kmin} 
\end{table}

Table \ref{tab:kmin} shows \(k_{\mathrm{min}}\) for different surveys.\footnote{The actual \(k_{\mathrm{min}}\) for DESI QSO is $ 0.0015 h \mathrm{Mpc}^{-1}$. However, we found the constraint on the turnover scale is slightly biased if we use that \(k_{\mathrm{min}}\). This is probably because we model the power spectrum on scales larger than the turnover as a quadratic function (see equation~(\ref{eq:P_dens})), while it should be approximately linear \citep{Harrison_1970, Zeldovich_1972}.} For DESI and 4HS, we set the maximum redshift of the peculiar velocity subsample to 0.15 \citep{Saulder_2023,Taylor_2023}. The \(k_{\mathrm{min}}\) of the velocity auto-power spectrum from DESI BGS and 4HS are only slightly smaller than \(k_{\mathrm{TO}}\), leaving it impossible for this spectrum to constrain the turnover scale alone. Therefore, we will only use the velocity auto-power spectrum when combined with other power spectra to fit the turnover scale.

\section{Cosmology from the turnover scale}
\label{sec:Theory}
\subsection{Converting the stretch parameter to cosmological parameters}
Ref.~\citep{Bahr_Kalus_2025} demonstrates with DESI mocks that defining a dilation parameter similar to analyses of Baryon Acoustic Oscillations,
\begin{equation}
    \alpha_{\mathrm{TO}} = \frac{k_{\mathrm{TO}}}{k_{\mathrm{TO}}^{\mathrm{fid}}} = \frac{D_V}{D_V^{\mathrm{fid}}} \frac{r_H^{\mathrm{fid}}}{r_H}
    \label{eq: alpha_TO_2_cosmo}
\end{equation}
works well to link our measurements of the turnover scale to cosmological parameters. This is required because our choice of fiducial cosmological model for converting spherical to Cartesian coordinates in estimating the power spectrum introduces a form of the Alcock-Paczynski effect \citep{Alcock_1979}. Here, the volume-averaged distance is given by 
\begin{equation}
    D_V(z) = \sqrt[3]{(1+z)^2 D_A(z)^2 D_H(z)},
    \label{eq: D_v}
\end{equation}
where \(D_A\) and \(D_H\) are the angular diameter and Hubble distances respectively.
%where the angular diameter distance \(D_A\) is defined as 
%\begin{equation}
%    D_A(z) = \frac{c}{(1+z)^2}\left(\int_0^{z} \frac{dz_1}{H_0E(z_1)}\right)^2,
%    \label{eq:D_A}
%\end{equation}
%the Hubble distance \(D_H\) is defined as 
%\begin{equation}
%    D_H(z) = \frac{cz}{H_0E(z)},
%    \label{eq:D_H}
%\end{equation}
%and finally 
%\begin{equation}
%    E(z) = \sqrt{\Omega_m(1+z)^3 + (1 - \Omega_m)}
%    \label{eq: Ez}
%\end{equation}
%for the \(\Lambda\)CDM cosmology. 
The other relevant quantity, the particle horizon size at matter-radiation equality, is given by 
\begin{equation}
    r_H = \frac{2c(\sqrt{2} - 1)\sqrt{a_{eq}}}{H_0\sqrt{\Omega_m}},
    \label{eq: r_H}
\end{equation}
where \(\Omega_m\) is the total matter density and \(a_{eq} = \frac{\Omega_r}{\Omega_m}\) is the scale factor at matter radiation equality. In equation~(\ref{eq: alpha_TO_2_cosmo}), the superscript `fid' represents these quantities for our fiducial cosmological model. To convert the stretch parameter \(\alpha_{\mathrm{TO}}\) to cosmological parameters, we can first jointly constrain the parameters \(r_H\) and \(\Omega_m\) from the posterior of \(\alpha_{\mathrm{TO}}\). This method is similar to the BAO analysis, where the dilation parameter \(\alpha\) is converted into a constraint on the sound horizon at the drag epoch \(r_d\) and \(\Omega_m\). This conversion helps verify the internal consistency across different redshift bins. Once we are satisfied that the constraints from different redshift bins are consistent, we can combine them to break the degeneracy between \(r_H\) and \(\Omega_m\) to further constrain the underlying cosmological parameters. 

To see this more clearly, we follow ref.~\citep{Komatsu_2009}, which shows the radiation density \(\Omega_r\) is related to the photon density \(\Omega_\gamma\) by
\begin{equation}
    \Omega_r = \Omega_\gamma(1 + 0.2271N_{\mathrm{eff}}),
    \label{eq: Omega_r}
\end{equation}
where \(N_{\mathrm{eff}}\) is the effective number of relativistic species. Assuming standard physics for radiation, the photon density is given by 
\begin{equation}
    \Omega_\gamma = \frac{8\pi G}{3H_0^2}\frac{4\sigma_BT_{\mathrm{CMB}}^4}{c^3},
    \label{eq: omega_gamma}
\end{equation}
where \(G\) is the gravitational constant, \(\sigma_B\) is the Stefan-Boltzmann constant, and \(c\) is the speed of light. Assuming \(N_{\mathrm{eff}}\) and \(T_{\mathrm{CMB}}\) are constants, then the radiation density is inversely proportional to \(H_0^2\). If we further assume a $\Lambda$CDM cosmological model then \(D_A\) and \(D_H\) depend only on $\Omega_{m}$ and $H_{0}$, leading to the result that the turnover scale and equation~(\ref{eq: alpha_TO_2_cosmo}) only depend on \(\Omega_m\) and \(H_0\). Substituting equations~(\ref{eq: D_v}), (\ref{eq: r_H}) and the relevant expression for the angular diameter and Hubble distances into equation~(\ref{eq: alpha_TO_2_cosmo}), results in 
\begin{align}
    \alpha_{\mathrm{TO}} &= \frac{H_0\sqrt{\Omega_m}}{2(\sqrt{2} - 1)\sqrt{B}}\sqrt[3]{\left(\int_0^{z} \frac{dz_1}{\mathcal{E}(z_1)}\right)^2 \frac{z}{\mathcal{E}(z)}} \frac{r_H^{\mathrm{fid}}}{D_V^{\mathrm{fid}}}
\end{align}
after simplification. Here, 
\begin{equation}
    B = \frac{8\pi G}{3}\frac{4\sigma_BT_{\mathrm{CMB}}^4}{c^3}(1 + 0.2271N_{\mathrm{eff}})
    \label{eq: factor_B}
\end{equation}
and 
\begin{equation}
    \mathcal{E}(z) = \sqrt{(1+z)^3 + \frac{1}{\Omega_m} -1}
    \label{eq: mathcal_E}. 
\end{equation}
This set of equations shows that, for fixed $N_{\mathrm{eff}}$, a single constraint on the turnover scale can probe a particular (redshift-dependent) degeneracy direction in the $\Omega_{m}-H_{0}$ plane, or equivalently provide an enclosed bound on the physical matter density $w_{m}=\Omega_{m}h^{2}$. In this work, we follow ref.~\citep{Bahr_Kalus_2023} to fix \(N_{\mathrm{eff}} = 3.046\) \citep{Mangano_2005} and \(T_{\mathrm{CMB}} = 2.72548\) K \citep{Fixsen_2009} when constraining \(\Omega_m\) and \(H_0\).

To extract cosmological constraints from the turnover scale in the matter power spectrum, we need a cosmology-independent way to identify the location of this peak. For this, we follow refs.~\citep{Bahr_Kalus_2023, Poole_2013, Blake_2005} to model the galaxy power spectrum around the turnover scale (\(k_{\mathrm{TO}}\)) using a broken power law, 
\begin{equation}
    P_{gg}(k) = \left\{\begin{array}{rcl} A_{gg}^{1-mx^2} & \mbox{for}
& x \leq 1 \\ A_{gg}^{1-n_{gg}x^2} & \mbox{for} & x > 1, 
\end{array}\right.
\label{eq:P_dens}
\end{equation}
with \(x = \log_{k_{\mathrm{TO}}}(k) - 1\). Both the wavenumber \(k\) and the turnover scale \(k_{\mathrm{TO}}\) are in units of \(h\,\mathrm{Mpc}^{-1}\). \(A_{gg}\) denotes the amplitude of the galaxy power spectrum \(P_{gg}\) at \(k = k_{\mathrm{TO}}\), \(m\) denotes the slope of the power spectrum at scales larger than the turnover scale and \(n\) denotes the slope at scales smaller than the turnover scale. Equation~(\ref{eq:P_dens}) is independent of cosmological models since we only use the fact that we can model any continuously differentiable function near the local extremum with a quadratic function. Based on linear theory, equation~(\ref{eq:P_dens}) also applies for the temperature power spectrum and the cross-power spectrum between temperature and galaxy density. To distinguish the amplitude and slope parameters from different power spectra, we use \(A_{TT}\) and \(A_{gT}\) to denote the amplitude of the temperature auto-power spectrum and the amplitude of the temperature-galaxy cross-power spectrum. Similarly, we also use \(n_{TT}\) and \(n_{gT}\) to denote the slope of the two power spectra on small scales, respectively. Note that we assume, based on linear theory, that all power spectra probe the same turnover scale and large-scale slope $m$. The differences in amplitude and small-scale slopes are to account for the different ``Kaiser''-like factors and non-linear damping we have in our mock data, but would also be expected in real data. 

\subsection{New model for the velocity power spectra}
Our extension to previous work involves using combined constraints from the density and velocity power spectra. As such, we need similarly cosmology-independent models for the turnover scale in both the density-velocity cross-power spectrum $P_{gv}$ and the velocity auto-power spectrum, $P_{vv}$. We build on equation~(\ref{eq:P_dens}) to model these with
\begin{equation}
    P_{\mathrm{gv}}(k) = \left\{ \begin{array}{rcl}
A_{\mathrm{gv}}^{1-mx^2}/k & \mbox{for}
& x \leq 1 \\ A_{\mathrm{gv}}^{1-n_{gv}x^2}/k & \mbox{for} & x > 1, 
\end{array}\right.
\label{eq:P_cross}
\end{equation}
and
\begin{equation}
    P_{\mathrm{vv}}(k) = \left\{ \begin{array}{rcl}
A_{\mathrm{vv}}^{1-mx^2}/k^2 & \mbox{for}
& x \leq 1 \\ A_{\mathrm{vv}}^{1-n_{vv}x^2}/k^2 & \mbox{for} & x > 1, 
\end{array}\right.
\label{eq:P_vel}
\end{equation}
respectively. Our models are motivated as follows: on large-scales, linear perturbation theory predicts \(P_{mm} = P_{\theta \theta} = P_{m \theta}\), with the conversion from velocity divergence to line-of-sight velocity introducing a factor of \(\frac{1}{k}\) for each velocity used in constructing the relevant power spectrum. Other than this, we use the same \(m\) as equation~(\ref{eq:P_dens}) to denote the slope of the power spectrum on scales larger than the turnover scale, as the three spectra should share the same underlying linear matter power spectrum. There is a difference in the amplitude at the turnover scale due to the large-scale RSD factors differing between the three spectra in linear theory (due to their different combinations of the galaxy bias and growth rate). Our models account for this difference, while remaining cosmology-independent, using different amplitude parameters for each power spectrum. 

On large scales, we expect different damping terms \(D_g\), \(D_u\), and \(D_T\) have no effect on the amplitude of the power spectrum at the turnover scale since \(k_{\mathrm{TO}}\mu \sigma_g \sim k_{\mathrm{TO}}\sigma_u \sim k_{\mathrm{TO}}\mu \sigma_{NL} \sim 0\) and \(D_g \approx D_u \approx D_T \sim 1\) based on previous measured values for \(\sigma_g\), \(\sigma_u\), and \(\sigma_{NL}\) \citep{Koda_2014, Li_2007}. However, the slope of our ``mock'' data power spectrum will differ significantly on small scales, as the damping factors are different. The mode de-projection we use to remove the imprint of Baryon Acoustic Oscillations in our analysis (see the next section, \ref{sec:Mode}) will down-weight scales smaller than the turnover scale, and suppress the impact of differences among the damping factors. We examine the effect of assuming the same slope on small scales when performing multi-tracer analysis on the turnover scale constraints. 

\begin{comment}
Although on quasi-nonlinear scales, \(P_{mm}, P_{\theta \theta},\) and \(P_{m \theta}\) are no longer the same, and the three measured spectra undergo different levels of non-linear RSD damping, (which is our "mock" dataset is dictated by their different combinations of \(D_u\) and \(D_g\)) none of this significantly affects the ability of our simple models to recover the turnover scale. Ref. \citep{Koda_2014} suggest both \(\sigma_g \approx 5\) and \(\sigma_u \approx 15\) well-represent galaxies found in different subhalo mass ranges and at different number densities, so \(k_{\mathrm{TO}}\mu \sigma_g \sim k_{\mathrm{TO}}\sigma_u \sim 0\) and \(D_g \approx D_u \approx 1\). Therefore, we do not expect non-linearities (and especially not those included in our mock data) to significantly affect the amplitude of the power spectrum at the turnover scale. Furthermore, the mode de-projection we use to remove the imprint of Baryon Acoustic Oscillations in our analysis (see section \ref{sec:Mode}) down-weights scales smaller than the turnover scale, further supressing the impact of differences among \(P_{mm}, P_{\theta \theta},\) and \(P_{m \theta}\) and between \(D_u\) and \(D_g\) on nonlinear scales
\end{comment}

\subsection{Model fitting}
\label{sec:modelfitting}
Given our data  and associated assumptions regarding their overlap and fitting scales, alongside our cosmology-independent models, we can write the likelihood for a multi-tracer analysis covering a single hemisphere as  
\begin{equation}
    \ln{L} \propto -\frac{1}{2}(\Delta\boldsymbol{p}^TC^{-1} \Delta\boldsymbol{p}).
\end{equation}
Here, \(\Delta\boldsymbol{p}\) represents the difference between the combined data vector \(\boldsymbol{p_d}\) and the combined model power spectra (\(\boldsymbol{p}\); i.e., the concatenated combination of \(P_{aa}\), \(P_{bb}\) and \(P_{ab}\)) for our analysis after the Gaussianisation process \citep{Bahr_Kalus_2023}
\begin{equation}
    \Delta\boldsymbol{p} = 3\boldsymbol{p_d}\left(1-\sqrt[3]{\frac{\boldsymbol{p}}{\boldsymbol{p_d}}}\right).
\end{equation}
We also combine the individual covariance matrices given in equation~\ref{eq:cov_full} above to generate the full covariance matrix \(C\) as
\begin{equation}
    C = \left(\begin{array}{ccc}
        C^{\mathrm{aaaa}}_{l_1l_2} & C^{\mathrm{aabb}}_{l_1l_2} & C^{\mathrm{aaab}}_{l_1l_2} \\
        C^{\mathrm{bbaa}}_{l_1l_2} & C^{\mathrm{bbbb}}_{l_1l_2} & C^{\mathrm{bbab}}_{l_1l_2} \\
        C^{\mathrm{abaa}}_{l_1l_2} & C^{\mathrm{abbb}}_{l_1l_2} & C^{\mathrm{abab}}_{l_1l_2}
    \end{array} \right).
    \label{eq:C_combined}
\end{equation}
For simplicity, we consider only the lowest non-zero multipoles of their respective power spectra as they contain the most information (\(l = 1\) for the galaxy-velocity cross-power spectrum and \(l = 0\) for all other power spectra).  We set \(k_{\mathrm{max}} = 0.20h\mathrm{Mpc}^{-1}\) because our quasi-linear model for the mock data power spectrum breaks down beyond \(k_{\mathrm{max}} = 0.20h\mathrm{Mpc}^{-1}\) \citep{Koda_2014} and ref.~\citep{Bahr_Kalus_2023, Bahr_Kalus_2025} also set \(k_{\mathrm{max}} = 0.20h\mathrm{Mpc}^{-1}\) when they are fitting real data. Future work could investigate whether including higher multipoles can improve the constraints on the turnover scale. 

Lastly, we can consider the combination of surveys across both the northern and southern hemispheres. As we treat these as independent, we can simply combine the MCMC chains of both surveys. We assume \(k_{\mathrm{TO}}\) and \(m\) will be the same from both surveys because \(P(k, z) = D(z)P(k, z = 0)\) in linear theory. Therefore, the slope of the power spectrum on large scales and the location of the turnover should remain the same. However, to account for potentially differing galaxy samples in each hemisphere, we keep different amplitude and small-scale slope parameters.

\subsection{Mode Deprojection}
\label{sec:Mode}
Ref. \citep{Bahr_Kalus_2023} found that the presence of BAO features could bias the constraints on the turnover scale. To mitigate this effect, they treat the BAO features as systematic contamination and down-weight the relevant scales with mode de-projection \citep{Rybicki_1992}. Here, we follow the same methodology, applied to our full set of three power spectra instead. We first calculate the power spectrum with the fiducial cosmological parameters. Then, we fit our models from the previous section to the mock power spectrum. The difference between the best-fit and the mock power spectrum is used to calculate the mode-deprojection. Lastly, we modified the covariance matrix using mode-deprojection to inflate the uncertainty on scales most affected by the BAO. 

Ref.~\citep{Bahr_Kalus_2023} implements this by defining a template for the residual galaxy power spectrum 
\begin{equation}
    f_k^{\mathrm{BAO}} = \left\{ \begin{array}{rcl}
0 & \mbox{for}
& x_{\mathrm{fid}} < 1 \\ P_{gg}^{\mathrm{fid}}(k) - P_{gg, \mathrm{BF}}^{1-n_{\mathrm{BF}} x_{\mathrm{fid}}^2} & \mbox{for} & x_{\mathrm{fid}} \geq 1, 
\end{array}\right.
\label{eq:f_k}
\end{equation}
where \(x_{\mathrm{fid}} = \log_{k^{\mathrm{fid}}_{\mathrm{TO}}}(k) - 1\) and \(k^{\mathrm{fid}}_{\mathrm{TO}}\), \(P^{\mathrm{fid}}_{\mathrm{gg}}\) are the turnover scale with the fiducial cosmology and the fiducial galaxy power spectrum. For our trio of power spectra, we modify equation~(\ref{eq:f_k}) to 
\begin{equation}
    f_k^{\mathrm{BAO}} = \left\{ \begin{array}{rcl}
0 & \mbox{for}
& x_{\mathrm{fid}} < 1 \\ \left( \begin{array}{c}
     P^{\mathrm{fid}}_{aa}(k)\\
     P^{\mathrm{fid}}_{bb}(k)\\
     P^{\mathrm{fid}}_{ab}(k)  
\end{array}\right) - \left(\begin{array}{c}
     P_{aa, \mathrm{BF}}^{1-n_{\mathrm{BF}}^{aa} x_{\mathrm{fid}}^2}\\
     P_{bb,\mathrm{BF}}^{1-n_{\mathrm{BF}}^{bb} x_{\mathrm{fid}}^2}/k^{2\gamma}\\
     P_{ab,\mathrm{BF}}^{1-n_{\mathrm{BF}}^{ab} x_{\mathrm{fid}}^2}/k^\gamma
\end{array}\right) & \mbox{for} & x_{\mathrm{fid}} \geq 1, 
\end{array}\right.
\label{eq:f_k_new}
\end{equation}
where \(P_{aa,\mathrm{BF}}\) and \(P_{bb,\mathrm{BF}}\) are the best-fit amplitudes of the auto-power spectrum for tracers \(a\) and \(b\), respectively. Similarly, \(P_{ab,\mathrm{BF}}\) is the best-fit amplitude of the cross-power spectrum. Additionally, \(n_{\mathrm{BF}}^{aa}\), \(n_{\mathrm{BF}}^{bb}\), and \(n_{\mathrm{BF}}^{ab}\) are the slopes for the auto-power spectrum of tracer \(a\), \(b\), and the slope for the cross-power spectrum, respectively. $\gamma=1$ if one of the tracers is the peculiar velocity field, and is zero otherwise. 

\begin{figure}
    \centering
    \includegraphics[width=0.328\textwidth]{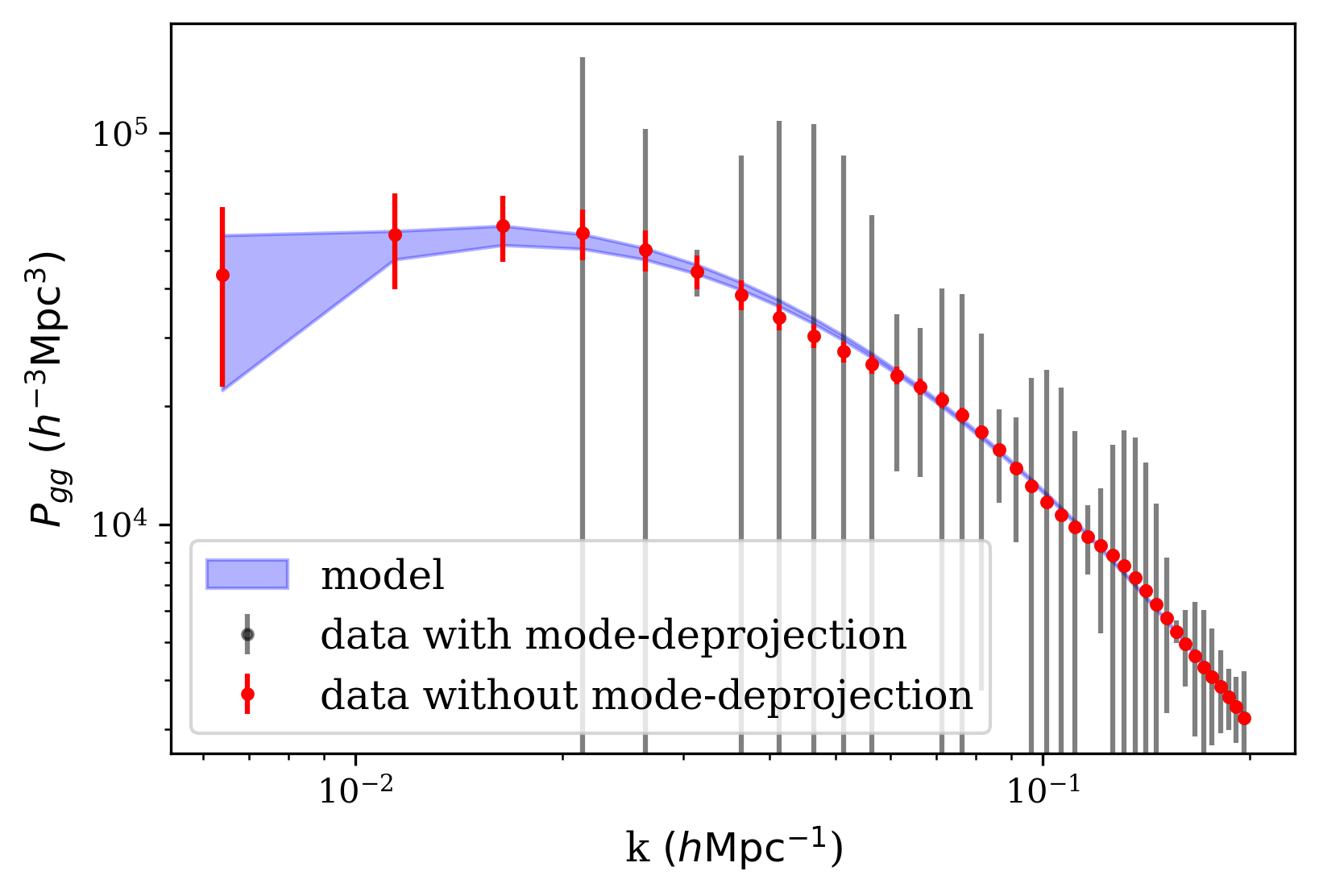}
    \includegraphics[width=0.328\textwidth]{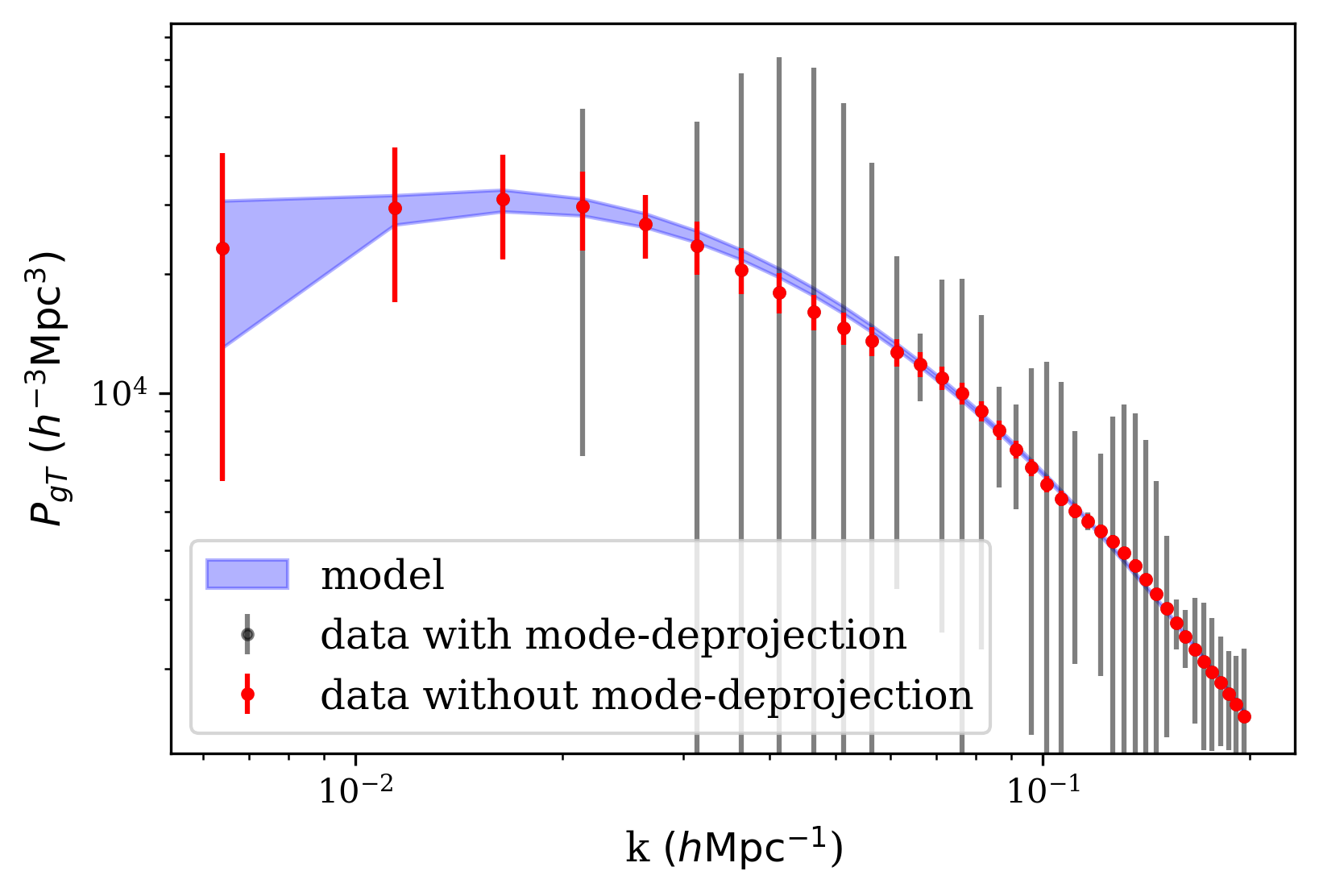}
    \includegraphics[width=0.328\textwidth]{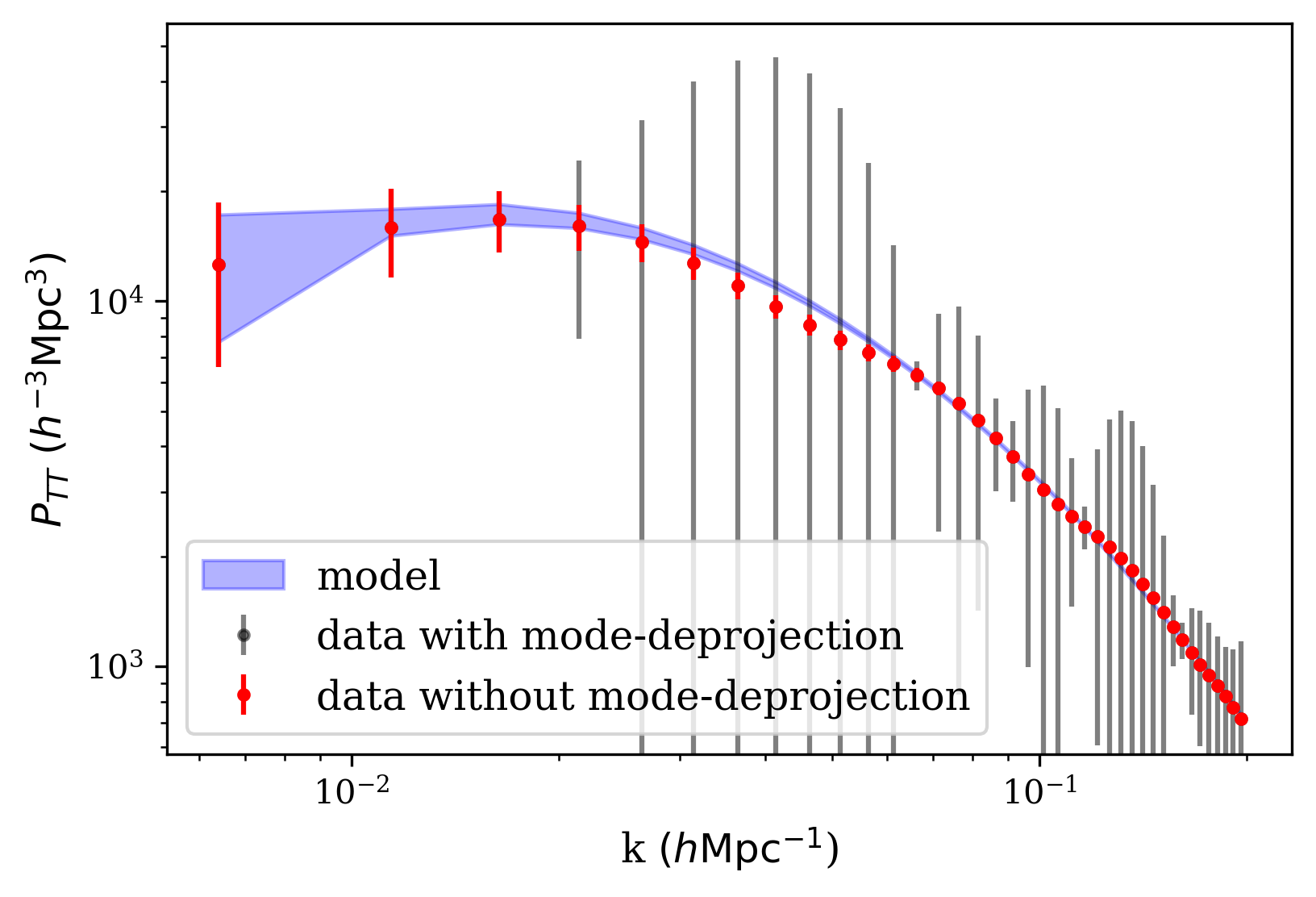}
    
    \caption{The blue band encloses the area of all possible model power spectra within \(1\sigma\) of the best-fit. The red error bars show the uncertainty of the mock power spectra before applying mode-deprojection, and the grey error bars are those after applying mode-deprojection. The mode-deprojection inflates the error bars on scales most affected by the BAO. We generate the galaxy power spectrum here for the 4HS survey and the temperature power spectrum for SKA1-B2. The model is generated with the best-fit parameters in Table.~\ref{tab:4HS_SKA_component}.}
    \label{fig:4HS_SKA1_PS_model}
\end{figure}

To find the various best-fit (BF) parameters, we first generate the linear power spectrum from CAMB \citep{Lewis_2000} with the fiducial cosmology. Then, we find the wavenumber at the maximum amplitude of the power spectrum and fit a quadratic function to the maximum and its two adjacent wavenumbers. We use the wavenumber at the maximum of the fitted quadratic function as \(k_{\mathrm{TO}}^{\mathrm{fid}}\). Then, we find the best-fit amplitude and slope of the power spectrum in equation~(\ref{eq:f_k_new}) using the \textsc{Basinhopping} \citep{Wales_1997} algorithm from \textsc{Scipy}. The best-fit parameters from the optimisation allow us to use equation~(\ref{eq:f_k_new}) to calculate \(f_k^{\mathrm{BAO}}\). The inverse covariance matrix \(\hat{C}\) after the mode de-projection is given by \citep{Bahr_Kalus_2023}
\begin{equation}
    \hat{C}^{-1} = C^{-1} - \frac{C^{-1}f_k^{\mathrm{BAO}}(f_k^{\mathrm{BAO}})^TC^{-1}}{(f_k^{\mathrm{BAO}})^TC^{-1}f_k^{\mathrm{BAO}}}
    \label{eq:inv_C}, 
\end{equation}
where \(C\) is the fiducial covariance matrix given by equation~(\ref{eq:C_combined}) and \(T\) is the transpose of a matrix. We then use the mode-deprojected inverse covariance matrix during the Markov Chain Monte Carlo (MCMC) analysis to find the constraints on the turnover scale. Fig.~\ref{fig:4HS_SKA1_PS_model} illustrates that the mode-deprojection massively inflates the uncertainty of the power spectrum at scales most affected by the BAO. 

\section{Results}
\label{sec:Results}
\subsection{Testing our method with the eBOSS QSO sample}
We begin by testing the applicability of our mock data analysis, which utilises an analytical covariance matrix, to real surveys. This benchmarking enables us to evaluate how well our forecasts will hold up in the real world. This also serves as a test to ensure that our pipelines can return unbiased constraints on the turnover scale. To do this, we generate a mock galaxy power spectrum using the fiducial cosmology in ref.~\citep{Bahr_Kalus_2023} for the eBOSS QSO sample. Ref.~\citep{Bahr_Kalus_2023} applied the same methodology to the eBOSS QSO data. Comparing our constraint on the turnover scale with that from ref.~\citep{Bahr_Kalus_2023} allows us to determine whether applying our analytical covariance matrix on the mock power spectrum can return a similar constraint on the turnover scale to that from applying the more accurate simulated covariance matrix to real data.

\begin{figure}
    \centering
	\includegraphics[width=0.70\textwidth]{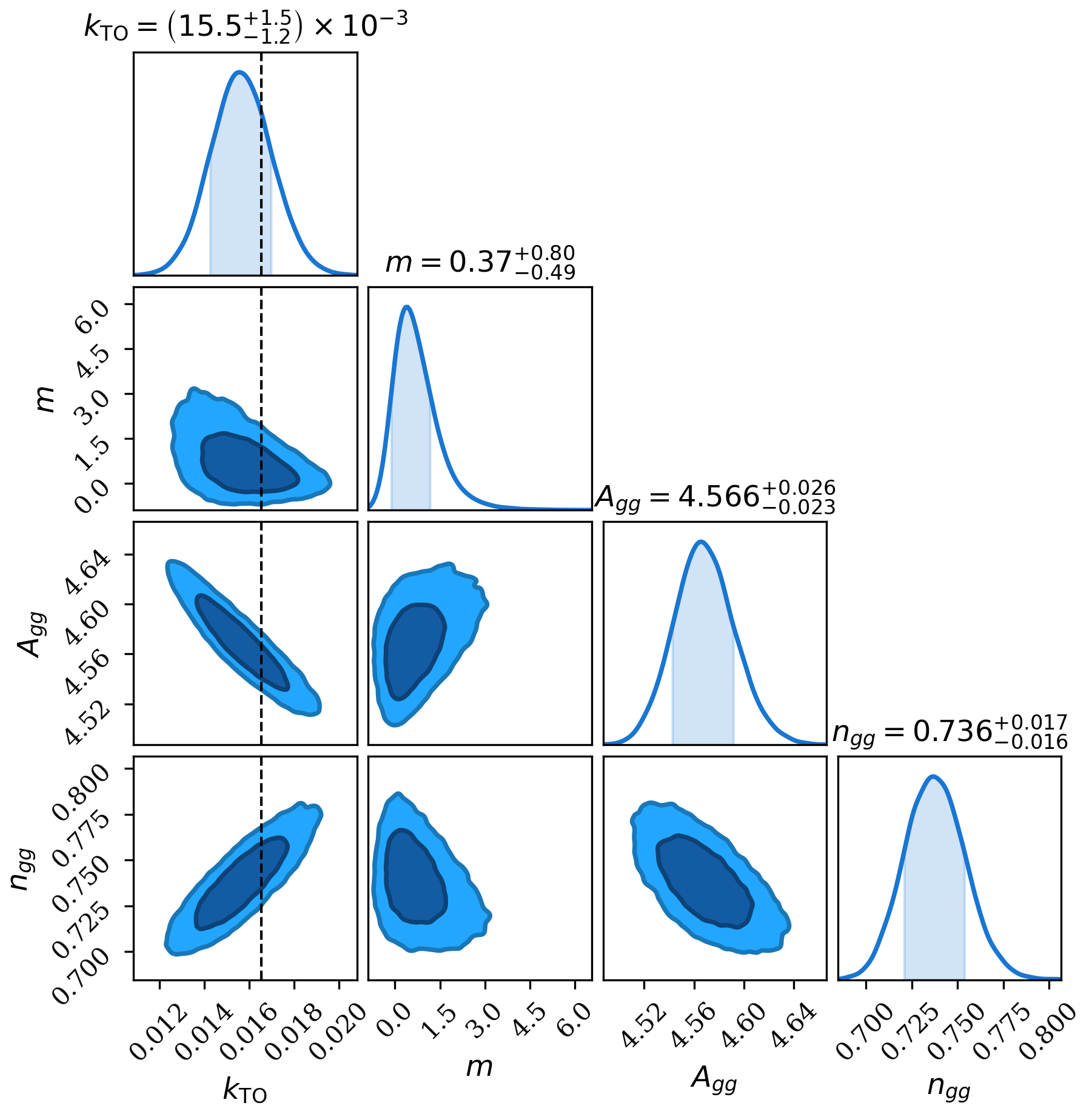}
    
    \caption{Constraints on the turnover scale with the fiducial cosmology and number density from ref.~\citep{Bahr_Kalus_2023}. The dashed line is the fiducial turnover scale. Our constraint is slightly tighter than the one in ref.~\citep{Bahr_Kalus_2023} because we do not have a window function except accounting for the binning effect, and the off-diagonal terms in the covariance matrix are neglected.}
    \label{fig:eBOSS}
\end{figure}

Our analytical covariance matrix is calculated using both the number density and the survey area from ref.~\citep{Bahr_Kalus_2023}. Fig.~\ref{fig:eBOSS} shows the corresponding constraints on the turnover scale \(k_{\mathrm{TO}} = 15.5_{-1.2}^{+1.5} \times 10^{-3} h \mathrm{Mpc}^{-1}\) is consistent with \(k_{\mathrm{TO}} = 17.6_{-1.9}^{+1.8} \times 10^{-3} h \mathrm{Mpc}^{-1}\) from ref.~\citep{Bahr_Kalus_2023}.\footnote{The mean of the posterior of \(k_{\mathrm{TO}}\) from ref.~\citep{Bahr_Kalus_2023} is around \(1\sigma\) away from ours. This is likely because the true cosmology of eBOSS differs from the fiducial cosmology in ref.~\citep{Bahr_Kalus_2023}. Using the Ez mocks \citep{Chuang_2014, Zhao_2021} with the fiducial cosmology, ref.~\citep{Bahr_Kalus_2023} got \(k_{\mathrm{TO}} = 16.7_{-1.9}^{+1.7} \times 10^{-3} h \mathrm{Mpc}^{-1}\), which is more consistent with our constraint.} The relative uncertainty (8.7\%) of our constraint on the turnover scale is slightly tighter than those (10.5\%) in ref.~\citep{Bahr_Kalus_2023}, which we attribute to the fact that our analysis does not incorporate the full survey window function, which will reduce the power somewhat on large scales even for $k > k_{\mathrm{min}}$. Fig.~3 in ref.~\citep{Bahr_Kalus_2023} demonstrates the impact the window function has on worsening the constraint on \(m\) and on \(k_{\mathrm{TO}}\). Additionally, ignoring the off-diagonal terms in the covariance matrix leads to an underestimation of the uncertainty. However, ref.~\citep{Wadekar_2020} shows that ignoring the off-diagonal terms only has a marginal (\(\lesssim 10\%\)) effect on the constraints on the cosmological parameters. Due to these simplifications, our analysis will yield a somewhat optimistic estimate of the turnover scale constraints for future surveys. However, as these same considerations affect both tracers in a multi-tracer analysis, we expect the relative improvement (or lack thereof) when using multiple tracers to be reliably predicted.  

\subsection{Forecasting the turnover scale constraint when combining galaxy density with peculiar velocity}

\begin{figure}
    \centering
	\includegraphics[width=0.70\textwidth]{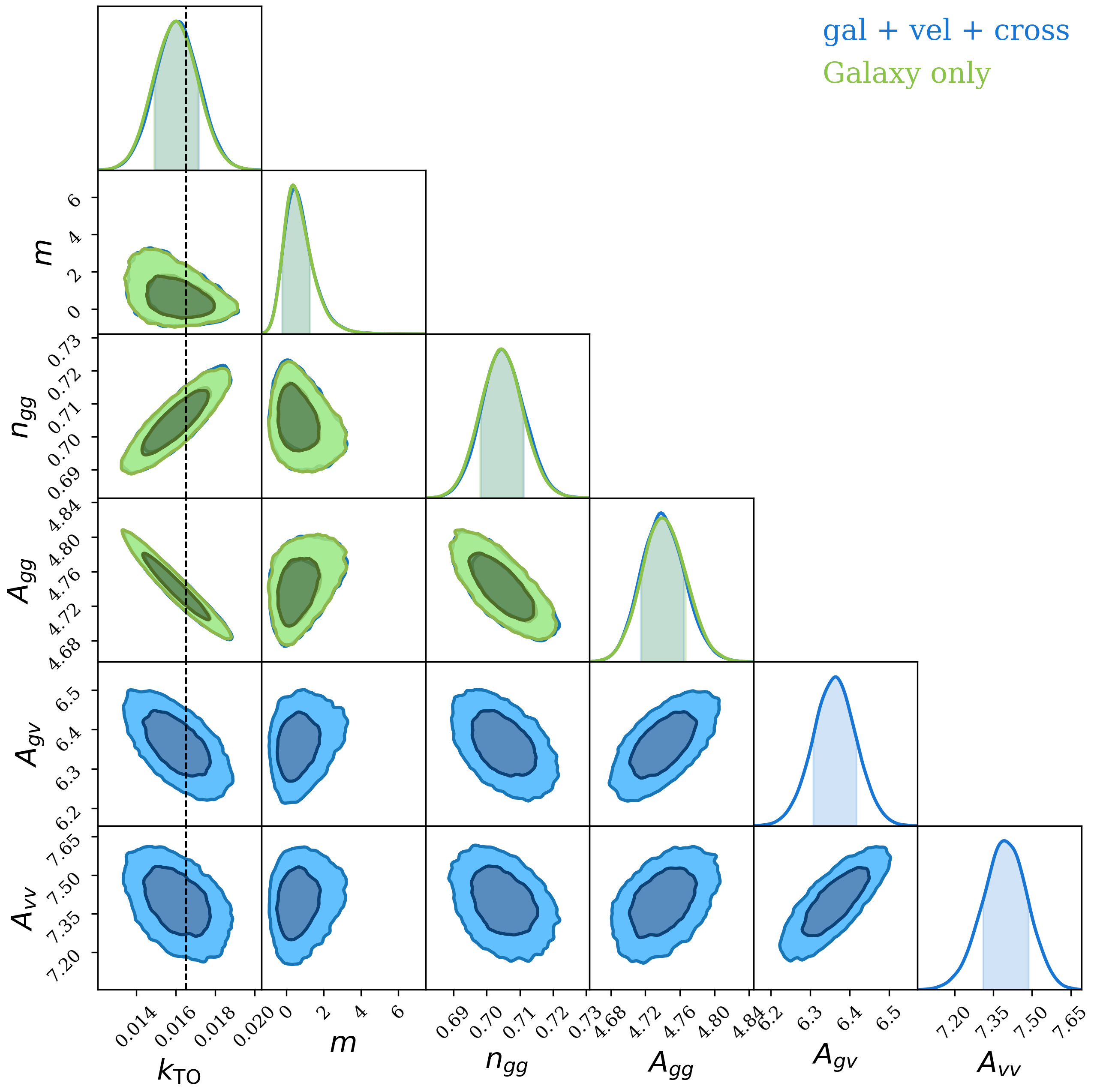}
    
    \caption{This figure illustrates the constraints on the turnover scale with only galaxy density (green) and combining the auto-power spectrum of galaxy density and peculiar velocity, and their cross-power spectrum (blue). When combining the galaxy density and peculiar velocity field, we also assume the small-scale slope \(n_{gg} = n_{gv} = n_{vv}\). We found that this assumption does not significantly impact the constraint on the turnover scale because most of the constraining power comes from the galaxy power spectrum. We found no noticeable improvement on the turnover scale constraint or in the detection probability of the turnover. For the first four parameters, the blue contours are the same as the green contours. }
    \label{fig:DESI_component}
\end{figure}

Fig.~\ref{fig:DESI_component} illustrates the constraints on the turnover scale with the DESI BGS + PV surveys. For the blue contours, we combine both the galaxy and peculiar velocity fields, while the green contours represent the constraints obtained by using only the galaxy density field. We also assume the small-scale slope \(n = n_{gg} = n_{gv} = n_{vv}\) when combining galaxy and peculiar velocity, and find it does not introduce any bias or improve on the turnover scale constraints. This is because the peculiar velocity has much less constraining power than the galaxy density. In Table.~\ref{tab:DESI_component}, we show the constraints on the free parameters, where the values inside the brackets are the best-fit. We also quote the best-fit parameters here because our marginalised posteriors for some surveys are affected by the projection effects \citep{Simon_2023}.\footnote{Evidence for this can be seen in Tables~\ref{tab:DESI_component}, ~\ref{tab:4HS_SKA_component}, ~\ref{tab:DESI_4HS_SKA1_component}, ~\ref{tab:DESI_4HS_SKA2_component}, where we see that tighter constraints on the turnover scale lead to better agreement between the mean of the posterior and the best-fit, reducing the projection effect.} For the 4HS survey, we also find that combining the galaxy and peculiar velocity field does not improve the constraints on the turnover scale and the detection probability. To investigate whether this result is caused by the lower number density of the peculiar velocity survey or its limited volume, we instead combine the 4HS redshift survey and the deeper LSST J<19 peculiar velocity sample to create a peculiar velocity catalogue that extends up to the redshift of 0.5, matching the maximum redshift of the galaxy survey. However, Table.~\ref{tab:DESI_component} demonstrates that there is still no improvement even in this case. Therefore, we conclude that peculiar velocity surveys cannot improve the constraints on the turnover scale when combined with galaxy redshift surveys because, for all current or planned surveys, either the maximum redshift or the error-weighted number density of peculiar velocities is just too low compared to the galaxy number density.    

\begin{table}[t!]
% title of Table
     % is used to refer this table in the text
\centering                          % used for centering table
\renewcommand{\arraystretch}{1.4} % Default value: 1
\begin{tabular}{c|c|c|c|c}        % centered columns (4 columns)
%\hline\hline                 % inserts double horizontal lines    % table heading 
Tracer(s) & $k_{\mathrm{TO}}$ $(h\mathrm{Mpc}^{-1})$ & $P(m>0)$ & $\alpha_{\mathrm{TO}}$ & $z_{\mathrm{eff}}$\\ \hline \hline
DESI BGS $g$ &  $0.0160_{-0.0011}^{+0.0011}(0.0162)$ & 82\% & $0.968_{-0.067}^{+0.067}(0.980)$ & 0.350\\
\hline
DESI BGS + PV $g + v$  & $0.0161_{-0.0011}^{+0.0011}(0.0164)$ & 82\% & $0.974_{-0.067}^{+0.067}(0.992)$ & 0.346\\
\hline
4HS $g$ &  $0.0158_{-0.0016}^{+0.0015}(0.0162)$ & 78\% & $0.956_{-0.097}^{+0.091}(0.980)$ & 0.257\\
\hline
4HS + LSST J < 19 $g + v$  & $0.0157_{-0.0016}^{+0.0015}(0.0162)$ & 81\% & $0.950_{-0.097}^{+0.091}(0.980)$ & 0.252\\
\hline
\end{tabular}
\caption{The constraints on \(k_{\mathrm{TO}}\), \(\alpha_{TO}\), and the turnover detection probability with the DESI BGS, 4HS, and LSST J<19 samples when combining the galaxy and peculiar velocity fields.  We also quote the best-fit parameters inside the brackets. Both combinations show that combining galaxy density and peculiar velocity does not improve the constraints on the turnover scale or the detection probability of the turnover.}   
\label{tab:DESI_component} 
\end{table}

\subsection{Forecasting the turnover scale constraint when combining galaxy density with intensity mapping}
\begin{figure}
    \centering
	\includegraphics[width=0.70\textwidth]{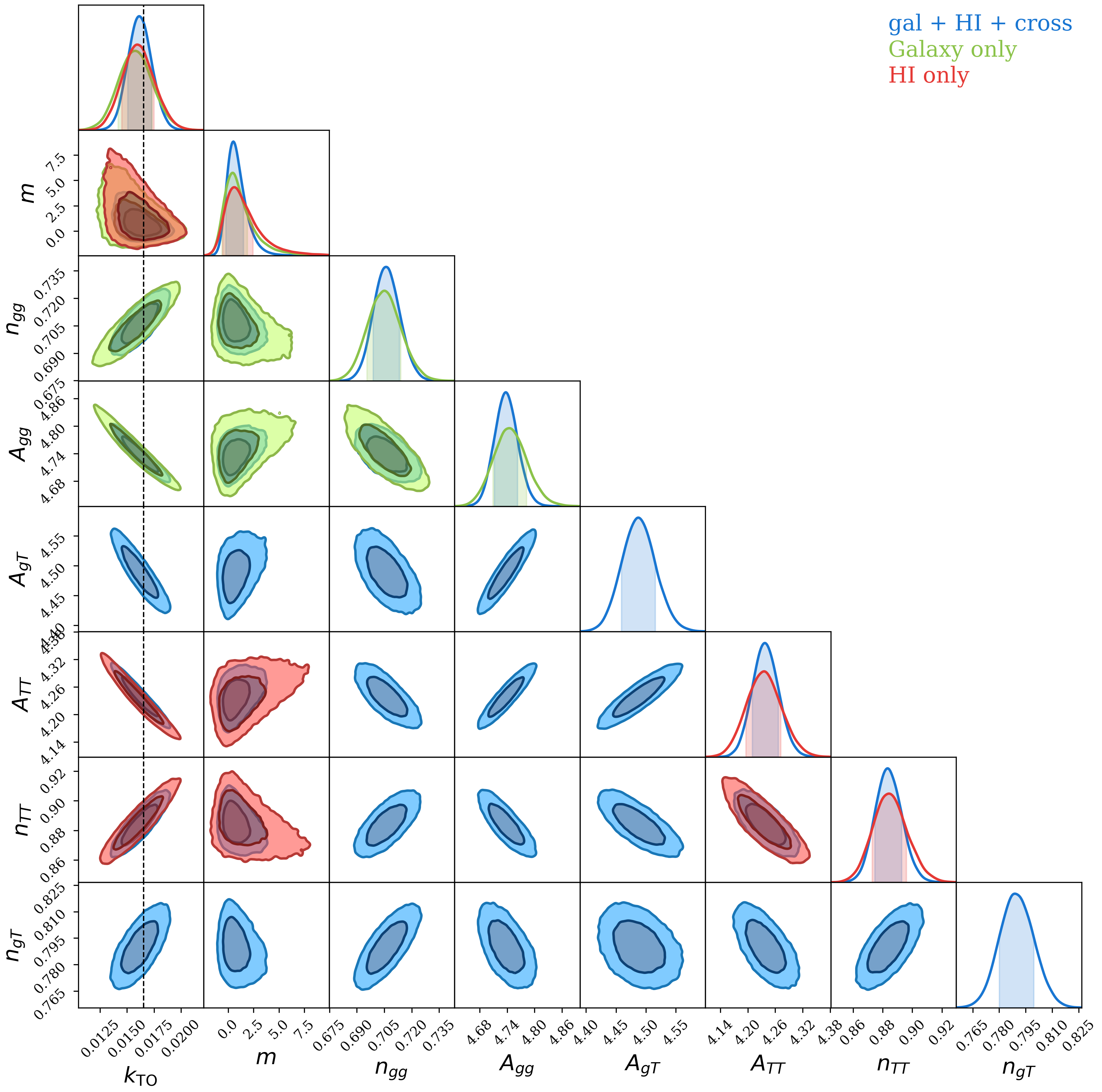}
    
    \caption{This figure illustrates the constraints on the turnover scale with the galaxy auto-power spectrum (green) from the 4HS survey, temperature auto-power spectrum (red) from the SKA1-B2 survey, and combining all three power spectra (blue). Different from Fig~\ref{fig:DESI_component}, combining galaxy density with intensity mapping improves the constraint on the turnover scale.}
    \label{fig:4HS_SKA_component}
\end{figure}

Our results so far indicate that the peculiar velocity field is unlikely to improve the constraints on the turnover scale. In this section, we investigate the combination of the galaxy redshift survey with HI intensity mapping. We first combined the 4HS galaxy redshift survey with the SKA1-B2 because they are in the same hemisphere. The constraint on the turnover scale is shown in Fig.~\ref{fig:4HS_SKA_component} and summarised in Table.~\ref{tab:4HS_SKA_component}. Unlike combining peculiar velocity with the galaxy density field, we find that we cannot assume the slopes of the galaxy power spectrum and the HI power spectrum are the same on small scales. Otherwise, our constraint on the turnover scale becomes biased. Unlike peculiar velocity, the HI power spectrum has a tight constraint on the slope \(n\). Fig.~\ref{fig:4HS_SKA_component} demonstrates \(n_{gg}\) and \(n_{TT}\) are more than \(6\sigma\) apart from each other, so we cannot assume the two to be the same. If only one tracer is used, both 4HS and SKA1-B2 return wider constraints than those from DESI BGS. For 4HS, this is because both its number density and maximum redshift are lower than those of DESI BGS. For SKA1-B2, its sky area is approximately one-third of that of DESI BGS. The covariance matrix is inversely proportional to the sky area, so the SKA1-B2 is expected to have less constraining power than DESI. However, unlike our finding for the combined DESI BGS + PV dataset, the combined constraint on the turnover scale is improved by around 31\% compared to using only 4HS and 27\% compared to using only SKA1-B2. The combined constraints on the turnover scale are then similar to those from DESI BGS. The detection probability of the turnover also increases by around 6\% for the combined sample compared to using only 4HS. We illustrate the best-fit power spectra model and the data power spectra before and after applying mode-deprojection in Fig.~\ref{fig:4HS_SKA1_PS_model}. The inverse covariance matrix after applying the mode-deprojection is almost singular. To obtain the error bars, we apply a small regularisation correction to the inverse covariance matrix 
\begin{equation}
    \hat{C}^{-1}_{R} = \hat{C}^{-1} + \epsilon I,
    \label{eq:regularisation}
\end{equation}
where \(\epsilon \ll \min(\mathrm{diag(\hat{C})})\) and \(I\) is the identity matrix. We chose \(\epsilon = 10^{-11} \sim 0.01\times \min(\mathrm{diag(\hat{C})})\) to generate Fig.~\ref{fig:4HS_SKA1_PS_model}.\footnote{The size of the error bar depends on \(\epsilon\). Nonetheless, the ``BAO wiggle'' pattern of the error bar remains. Since the \(\epsilon \ll \min(\mathrm{diag(\hat{C})})\), we will still get the same constraint on the turnover scale if we use the regularised inverse covariance matrix instead.} After applying the mode-deprojection, the pattern of the error bar has a similar shape to the BAO wiggle. This is expected since equation~(\ref{eq:inv_C}) demonstrates that scales that are most affected by the BAO will have the largest inflation of their error bars. Fig.~\ref{fig:4HS_SKA1_PS_model} also illustrates that the model is a good fit for the mock power spectra after applying the mode-deprojection.  

\begin{figure}
    \centering
	\includegraphics[width=1.0\textwidth]{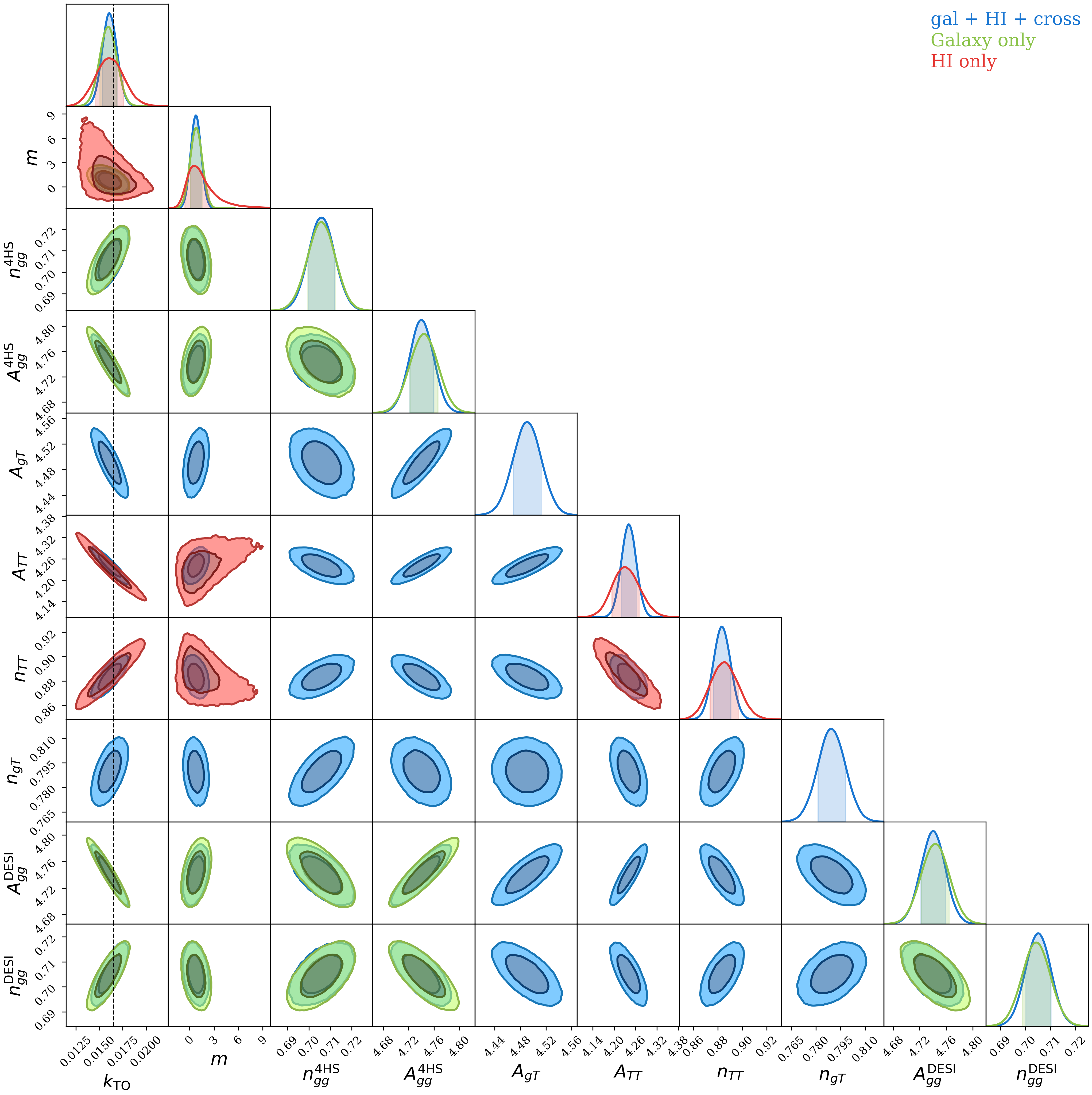}
    
    \caption{This figure illustrates the constraints on the turnover scale with the galaxy auto-power spectrum (green) from the combined 4HS and DESI survey, temperature auto-power spectrum (red) from the SKA1-B2 survey, and combining all three power spectra (blue). Similar to Fig.~\ref{fig:4HS_SKA_component}, combining the galaxy survey with the intensity mapping improves the turnover scale constraints. }
    \label{fig:DESI_4HS_SKA1_component}
\end{figure}

Next, we combine 4HS, SKA1-B2, and DESI BGS using the method/assumption of hemispherical independence outlined in section~\ref {sec:modelfitting}. We combined the chain of 4HS + SKA1-B2 and the chain of DESI. Therefore, we have different slope and amplitude parameters for the DESI power spectrum and the 4HS power spectrum in Fig.~\ref{fig:DESI_4HS_SKA1_component}. Additionally, the amplitude and slope parameters for the cross-power spectrum correspond to the cross-power spectrum between SKA1-B2 and 4HS, since we assume there is no overlap between DESI and SKA1-B2. The constraints on the free parameters are shown in Fig.~\ref{fig:DESI_4HS_SKA1_component} and summarised in Table.~\ref{tab:DESI_4HS_SKA1_component}. With only galaxy density, combining DESI BGS and 4HS improves the constraints on the turnover scale by 17\% compared to using only DESI BGS and 41\% compared to using only 4HS. After combining with SKA1-B2, the constraint on the turnover scale is improved by a further 14\% compared to using only the galaxy density. It also improves the detection probability of the turnover by around 3\%. The detection probability after combining 4HS, DESI BGS, and SKA1-B2 is similar to that from the DESI Y1 LRG sample \citep{Bahr_Kalus_2025}, which covers a much larger volume. Since DESI BGS, 4HS, and SKA1-B2 are either ongoing or planned to start in the next few years, we can potentially obtain a \(\sim5\%\) level constraint on the turnover scale and \(\sim90\%\) detection probability of the turnover using only redshift samples below 0.5 within the next decade. 

Lastly, we propose a hypothetical DESI++ catalogue, which covers the sky area of the combined DESI and 4HS survey (31000 \(\mathrm{deg}^2\)) and has the same density as the DESI BGS in both the northern and southern sky. In the northern hemisphere, this will already be achieved by the DESI survey. Although the 4HS survey in the southern hemisphere will not quite reach the DESI BGS number density for $z>0.1$, it could potentially be reached by an extension to 4HS carried out with the same 4MOST instrumentation after the current 5 year plan, a hypothetical J<19 galaxy sample \citep{Howlett_2017b}, or with the planned Widefield Spectroscopic Telescope \citep{Bacon_2024, Mainieri_2024}. 

\begin{figure}
    \centering
	\includegraphics[width=0.70\textwidth]{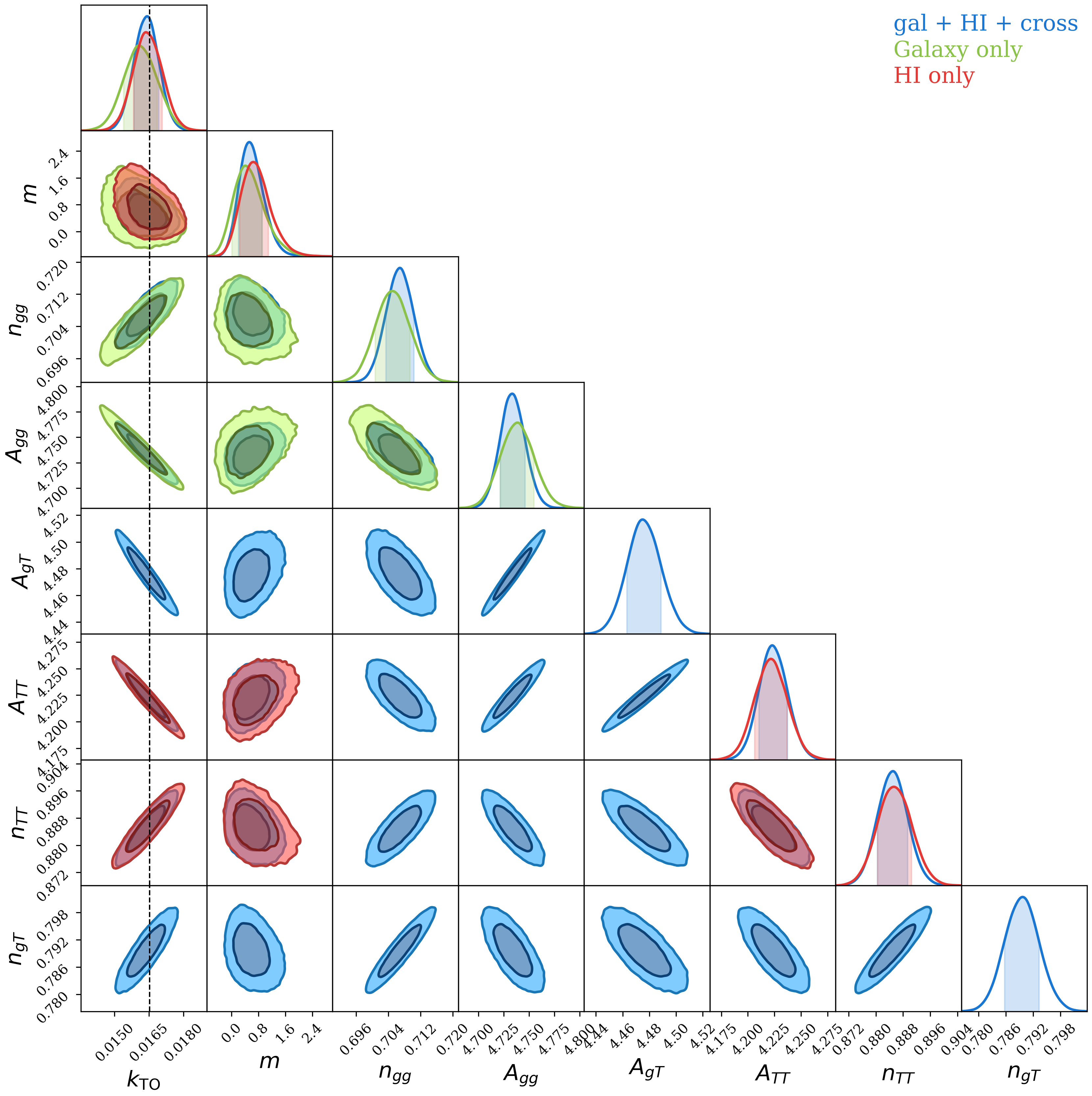}
    
    \caption{This figure illustrates the constraints on the turnover scale with the galaxy auto-power spectrum (green) from a hypothetical DESI++ catalogue (which has the sky area of the combined DESI and 4HS survey with the DESI number density), temperature auto-power spectrum (red) from the SKA2-B2 survey, and combining all three power spectra (blue). Similar to Fig.~\ref{fig:4HS_SKA_component} and \ref{fig:DESI_4HS_SKA1_component}, combining the galaxy survey with the intensity mapping improve the turnover scale constraints. }
    \label{fig:DESI_4HS_SKA2_component}
\end{figure}

Fig.~\ref{fig:DESI_4HS_SKA2_component} illustrates the constraint on the turnover scale by combining this hypothetical DESI++ catalogue with SKA2-B2. There is around a factor of \(\sqrt{6}\) improvement on the turnover scale constraints from SKA2-B2 compared to SKA1-B2. This is because we assume the configuration of SKA2-B2 to be the same as those from SKA1-B2, with a 6 times larger sky area. The covariance matrix is inversely proportional to the sky area, resulting in a \(\sqrt{6}\) improvement. Similarly, the DESI++ catalogue achieves a factor of \(\sqrt{2}\) improvement in the constraints on the turnover scale compared to DESI BGS due to its larger sky area. The combination of DESI++ with the SKA2-B2 improves the constraint on the turnover scale to around the 3\% level and the detection probability of the turnover to around 97\%. The constraining power of the combined sample is similar to that of the forecast constraint on the turnover scale from SKA1-B1 (Band-1, rather than the Band-2 survey used here) \citep{Cunnington_2022} and DESI QSO \citep{Bahr_Kalus_2023}, which both cover significantly larger volumes at higher redshift. The Widefield Spectroscopic Telescope and the SKA2-B2 are both planned to start around the 2040s.     

\begin{table}[t!]
% title of Table
     % is used to refer this table in the text
\centering                          % used for centering table
\renewcommand{\arraystretch}{1.4} % Default value: 1
\begin{tabular}{c|c|c|c|c}        % centered columns (4 columns)
%\hline\hline                 % inserts double horizontal lines    % table heading 
Tracer(s) & $k_{\mathrm{TO}}$ $(h\mathrm{Mpc}^{-1})$ & $P(m>0)$ & $\alpha_{\mathrm{TO}}$ & $z_{\mathrm{eff}}$\\ \hline \hline
4HS $g$ &  $0.0158_{-0.0016}^{+0.0015}(0.0162)$ & 78\% & $0.956_{-0.097}^{+0.091}(0.980)$ & 0.257\\
\hline
SKA1-B2 $T$ &  $0.0159_{-0.0014}^{+0.0016}(0.0166)$ & 82\% & $0.962_{-0.085}^{+0.097}(1.004)$ & 0.361\\
\hline
4HS $g$ + SKA1-B2 $T$  & $0.0161_{-0.0011}^{+0.0011}(0.0165)$ & 84\% & $0.974_{-0.067}^{+0.067}(0.998)$ & 0.303\\
\hline
\end{tabular}
\caption{The constraints on \(k_{\mathrm{TO}}\), \(\alpha_{TO}\), and the turnover detection probability with the 4HS survey and the SKA1-B2 survey. Combining the galaxy survey and the intensity mapping survey can improve the turnover scale by 31\% compared to using only the galaxy auto-power spectrum, and by 27\% compared to using only the temperature auto-power spectrum. Additionally, the probability of detecting the turnover also increases by around 6\% for the multi-tracer approach compared to only using the galaxy survey. }    
\label{tab:4HS_SKA_component} 
\end{table}

\begin{table}[t!]
% title of Table
     % is used to refer this table in the text
\centering                          % used for centering table
\renewcommand{\arraystretch}{1.4} % Default value: 1
\begin{tabular}{c|c|c|c|c}        % centered columns (4 columns)
%\hline\hline                 % inserts double horizontal lines    % table heading 
Tracer(s) & $k_{\mathrm{TO}}$ $(h\mathrm{Mpc}^{-1})$ & $P(m>0)$ & $\alpha_{\mathrm{TO}}$ & $z_{\mathrm{eff}}$\\ \hline \hline
DESI+4HS $g$ &  $0.01597_{-0.00093}^{+0.00090}(0.01606)$ & 87\% & $0.966_{-0.056}^{+0.054}(0.972)$ & N.A\\
\hline
SKA1-B2 $T$ &  $0.0159_{-0.0014}^{+0.0016}(0.0166)$ & 82\% & $0.962_{-0.085}^{+0.097}(1.004)$ & 0.361\\
\hline
DESI+4HS $g$ + SKA1-B2 $T$  & $0.01606_{-0.00075}^{+0.00082}(0.01602)$ & 89\% & $0.972_{-0.045}^{+0.050}(0.969)$ & N.A\\
\hline
\end{tabular}
\caption{The constraints on \(k_{\mathrm{TO}}\), \(\alpha_{TO}\), and the turnover detection probability with the combined DESI and 4HS survey and the SKA1-B2 survey. Combining the galaxy survey and the intensity mapping survey can improve the turnover scale by 14\% compared to using only the galaxy auto-power spectrum and 48\% compared to using only the temperature auto-power spectrum. Compared to Table.~\ref{tab:4HS_SKA_component}, the improvement in the turnover scale constraint compared to that with only the galaxy auto-power spectrum is weaker. This is because the galaxy auto-power spectrum from the combined DESI and 4HS survey has stronger constraining power than the SKA1-B2 survey.}    
\label{tab:DESI_4HS_SKA1_component} 
\end{table}

\begin{table}[t!]
% title of Table
     % is used to refer this table in the text
\centering                          % used for centering table
\renewcommand{\arraystretch}{1.4} % Default value: 1
\begin{tabular}{c|c|c|c|c}        % centered columns (4 columns)
%\hline\hline                 % inserts double horizontal lines    % table heading 
Tracer(s) & $k_{\mathrm{TO}}$ $(h\mathrm{Mpc}^{-1})$ & $P(m>0)$ & $\alpha_{\mathrm{TO}}$ & $z_{\mathrm{eff}}$\\ \hline \hline
DESI++ $g$ &  $0.01601_{-0.00062}^{+0.00084}(0.01622)$ & 90\% & $0.969_{-0.038}^{+0.051}(0.981)$ & 0.350\\
\hline
SKA2-B2 $T$ &  $0.01634_{-0.00052}^{+0.00073}(0.01651)$ & 97\% & $0.989_{-0.031}^{+0.044}(0.999)$ & 0.361\\
\hline
DESI++ $g$ + SKA2-B2 $T$  & $0.01642_{-0.00057}^{+0.00051}(0.01646)$ & 97\% & $0.993_{-0.034}^{+0.031}(0.996)$ & 0.356\\
\hline
\end{tabular}
\caption{The constraints on \(k_{\mathrm{TO}}\), \(\alpha_{TO}\), and the turnover detection probability with the hypothetical DESI++ catalogue and the SKA2-B2 survey. Combining the galaxy survey and the intensity mapping survey can improve the turnover scale by 26\% compared to using only the galaxy auto-power spectrum and 14\% compared to using only the temperature auto-power spectrum. %Different from Table.~\ref{tab:4HS_SKA_component}, the improvement in the turnover scale constraint compared to that with only the temperature auto-power spectrum is weaker because the SKA2-B2 survey has much stronger constraining power than the galaxy survey.
}    
\label{tab:DESI_4HS_SKA2_component} 
\end{table}

\subsection{Combining low-redshift turnover scale constraints with the high-redshift constraints}
\label{sec:combined}

Our results so far have investigated the potential of individual and multi-tracer surveys to constrain the turnover scale below $z=0.5$. We have found that the combination of redshift surveys from DESI and 4HS, as well as intensity mapping from SKA Phase 1 in particular, can lead to a $\sim5\%$ constraint on this scale. However, does such a constraint have value in constraining cosmology compared to similar or better constraints that may be obtained at higher redshifts? We answer this question in this section.

Ref.~\citep{Bahr_Kalus_2023} demonstrates that the constraints on \(H_0\) and \(\Omega_m\) from the stretch factor are highly degenerate. However, we can improve the constraints by combining constraints on the stretch factor from multiple redshift bins. In this section, we will combine the constraints on the stretch factor from the various low-redshift samples we have tested with the expected constraints from the DESI LRG, ELG, and QSO samples, again produced using our mock data production method. The number densities for LRG, ELG, and QSO used here are taken from ref.~\citep{DESI_2016a}. Although the redshift ranges of the three tracers overlap, we will treat them as independent tracers. This combination will provide upper bounds on the potential constraints on \(\Omega_m\) and \(H_0\) that can be obtained by combining high- and low-redshift measurements of the turnover scale.

\begin{figure}
    \centering
	\includegraphics[width=0.70\textwidth]{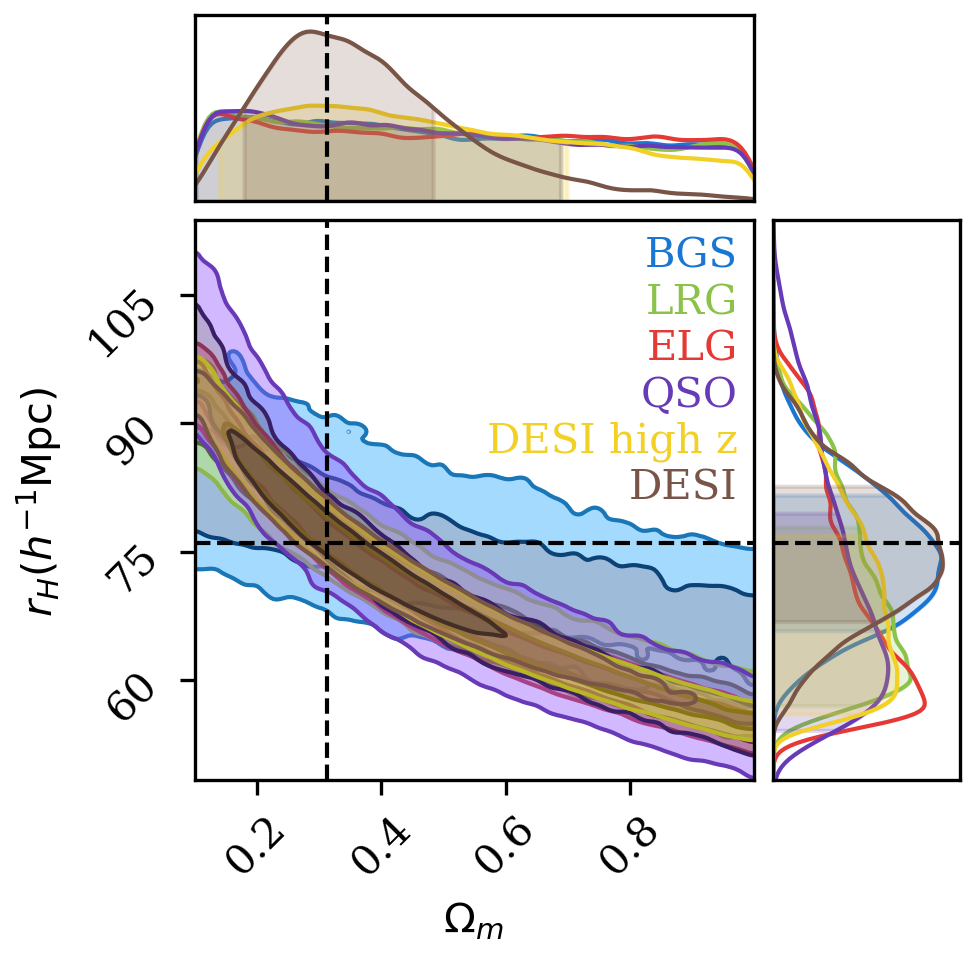}
    
    \caption{The constraints on \(r_H\) and \(\Omega_m\) from only galaxy redshift surveys: DESI BGS, LRG, ELG, QSO, and their combinations (``DESI high-z'' is LRG+ELG+QSO, while ``DESI'' is the combination of this with BGS). Similar to \citep{Bahr_Kalus_2023}, \(r_H\) and \(\Omega_m\) are highly degenerate when only using one redshift bin. However, constraints from different redshift bins overlap with each other and are consistent with the truth indicated by the black dashed line. Furthermore, the effective redshift for different galaxy samples differs, so the degeneracy direction between \(r_H\) and \(\Omega_m\) is also different. Therefore, combining different redshift bins will get tighter constraints on \(r_H\) and \(\Omega_m\).}
    \label{fig:DESI_rH_omegam}
\end{figure}

Fig.~\ref{fig:DESI_rH_omegam} illustrates the constraints on \(\Omega_m\) and \(r_H\) with different combinations of redshift bins. When fitting only a single redshift bin, \(\Omega_m\) and \(r_H\) are highly degenerate, as expected. However, they all have different effective redshifts, so the degeneracy directions for different redshift bins differ. Therefore, combining different redshift bins improves the constraints on \(\Omega_m\) and \(r_H\). The constraints on \(\Omega_m\) and \(r_H\) are still weak after combining the LRG, ELG, and QSO samples because they have similar effective redshifts (\(z_{\mathrm{eff}}^{\mathrm{LRG}} = 0.83, z_{\mathrm{eff}}^{\mathrm{LRG}} = 1.1, z_{\mathrm{eff}}^{\mathrm{LRG}} = 1.3\)) and the degeneracy directions are similar. However, the effective redshift for the BGS sample is 0.35 for the galaxy sample. This difference in the effective redshift results in the degeneracy direction of the BGS sample Fig.~\ref{fig:DESI_rH_omegam} being quite different from the high-redshift samples. Consequently, combining the BGS samples with LRG, ELG, and QSO can significantly reduce the uncertainty on \(\Omega_m\) and \(r_H\). Table.~\ref{tab:rH_omegam} demonstrates that by adding the BGS galaxy sample, we can improve the constraint on \(\Omega_m\) by 46\% and \(r_H\) by 25\% compared to only combining the LRG, ELG, and QSO samples. 

\begin{figure}
    \centering
    \includegraphics[width=0.495\textwidth]
    {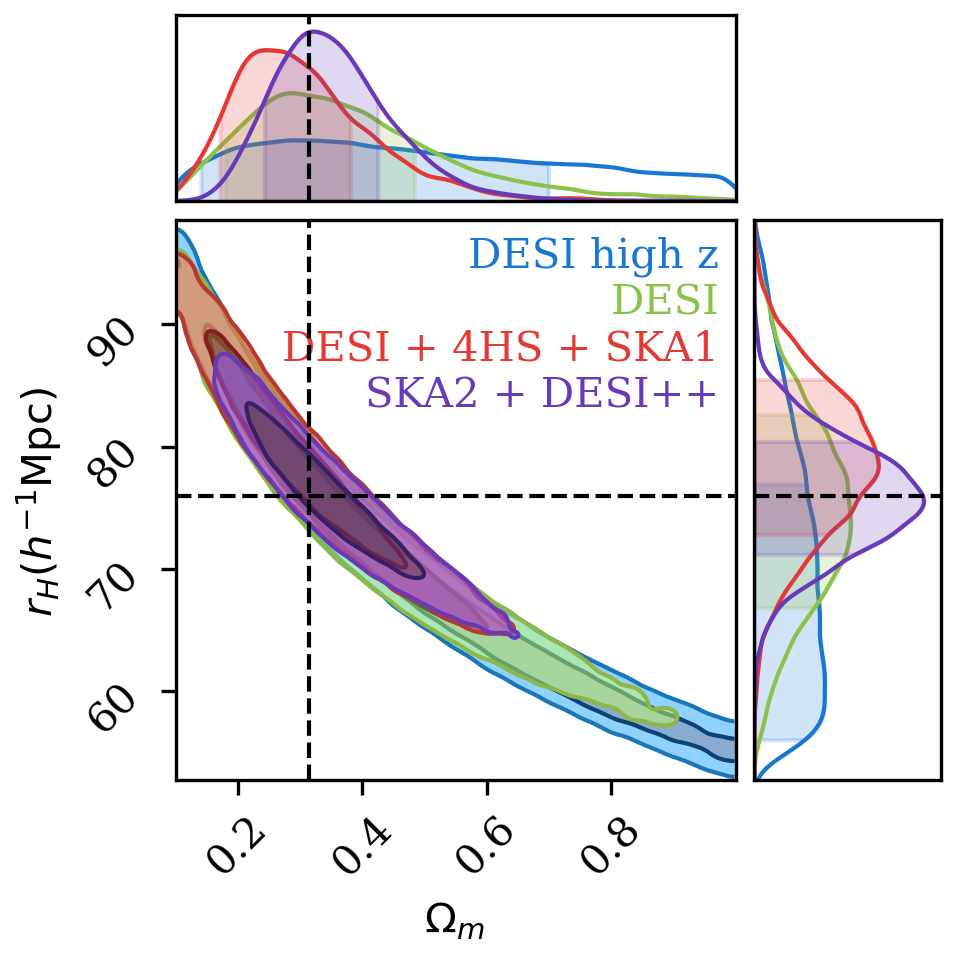}
    \includegraphics[width=0.495\textwidth]{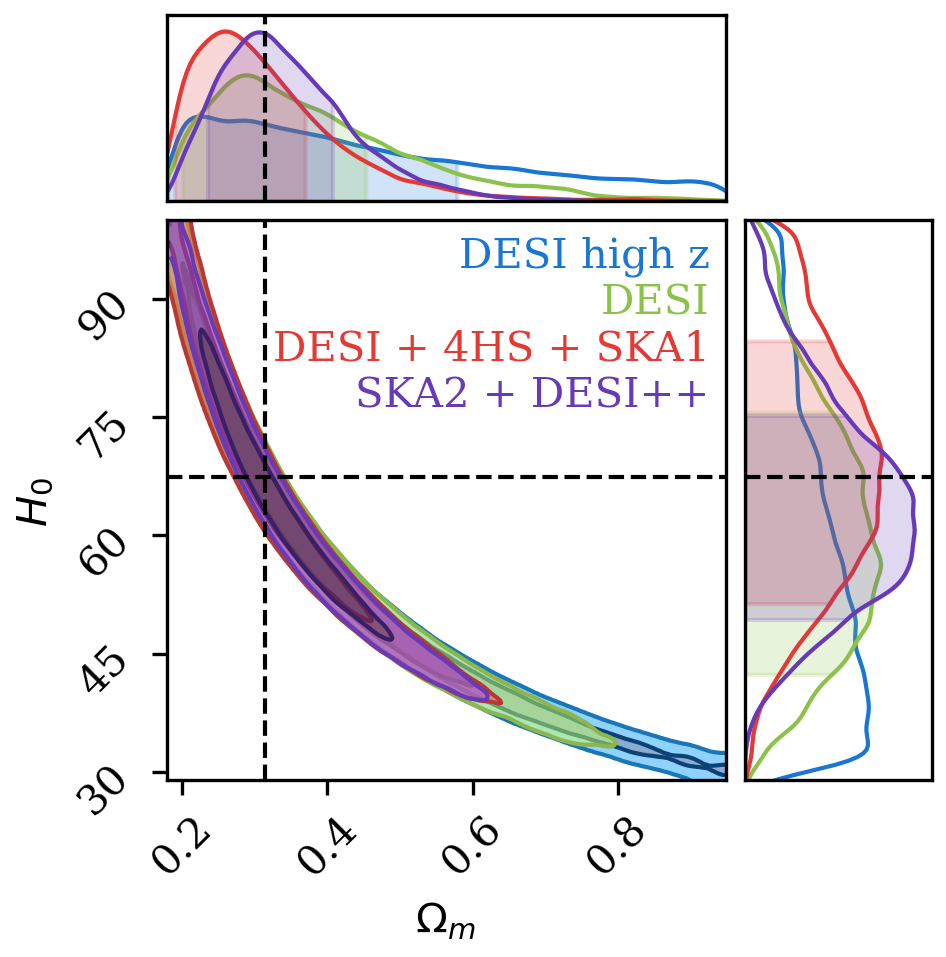}
    \caption{The constraints on \(r_H\) and \(\Omega_m\) (left), and \(H_0\) and \(\Omega_m\) (right) from DESI, DESI + 4HS + SKA1-B2, and SKA2-B2 + DESI++. Combining different high-redshift samples with a multi-tracer low-redshift sample significantly tightens the constraints on \(\Omega_m\) and \(r_H\). However, \(\Omega_m\) and \(H_0\) remain quite degenerate after combining multiple redshift bins and would require additional complementary datasets to disentangle.}
    \label{fig:rH_omegam}
\end{figure}

In addition to only the DESI BGS redshift sample, Fig~\ref{fig:rH_omegam} shows the constraints on \(r_H\) and \(\Omega_m\) when combining with other low-redshift probes as part of a multi-tracer analysis. Obviously, the constraints on both parameters are tighter if the low-redshift sample has a tighter constraint on the turnover scale. Table.~\ref{tab:rH_omegam} summarises the constraints on \(r_H\) and \(\Omega_m\) with different combinations. If we combine SKA1-B2 with 4HS and all DESI redshift bins, the constraints on \(\Omega_m\) and \(r_H\) are improved by a further 30\% and 20\%, respectively, compared to only with the DESI galaxy redshift surveys. Compared to constraints using only the high redshift samples, we see improvements of a factor of $\sim 2$ when adding a multi-tracer low-redshift analysis. With SKA2-B2 and the DESI++ catalogue, \(\Omega_m\) and \(r_H\) constraints are improved again by a further 14\% and 27\%, respectively compared to the data from 4HS, DESI and SKA1-B2. 

As a final note, although Fig.~\ref{fig:rH_omegam} clearly demonstrates that combining a low redshift multi-tracer measurement of the turnover scale with higher redshift measurements could improve the constraint on \(\Omega_m\) and \(H_0\), these two parameters are still degenerate, and we would need to combine with other probes to break the degeneracy. 

\begin{table}[t!]
% title of Table
     % is used to refer this table in the text
\centering                          % used for centering table
\renewcommand{\arraystretch}{1.4} % Default value: 1
\begin{tabular}{c|c|c}        % centered columns (4 columns)
%\hline\hline                 % inserts double horizontal lines    % table heading 
Data & $\Omega_m$ & $r_H$ ($h^{-1}\mathrm{Mpc}$) \\ \hline \hline
LRG + ELG + QSO (DESI high z)  &$0.28_{-0.14}^{+0.42}$ & $60.9_{-4.9}^{+16.0}$ \\
\hline
BGS + LRG + ELG + QSO (DESI) & $0.29_{-0.11}^{+0.19}$ & $73.1_{-6.3}^{+9.5} $ \\
\hline
4HS + SKA1-B2 + DESI  &$0.24_{-0.07}^{+0.14}$ & $78.2_{-5.4}^{+7.3}$ \\
\hline
SKA2-B2 + DESI++  &$0.31_{-0.07}^{+0.11}$ & $75.5_{-4.3}^{+5.0}$ \\
\hline
\end{tabular}
\caption{The constraints on \(\Omega_m\) and \(r_H\) after combining multiple redshift bins and different surveys. If we combine SKA1-B2 with 4HS and all DESI redshift bins, the constraint on \(\Omega_m\) is improved by 30\% and the constraint on \(r_H\) is improved by 20\% compared to only with the DESI redshift bins. With SKA2-B2 and the DESI++ catalogue, \(\Omega_m\) and \(r_H\) constraints are further improve by 14\% and 27\%, respectively. }    
\label{tab:rH_omegam} 
\end{table}

\section{Conclusion}
\label{sec:Conclusion}
In this work, we have investigated the use of multi-tracer techniques, involving a galaxy redshift survey combined with either a peculiar velocity or HI intensity mapping sample, to constrain the turnover scale in the matter power spectrum at $z < 0.5$. We have then demonstrated the utility of combining low-redshift measurements with the anticipated $z>0.5$ measurements from DESI. 

To achieve this, we have generated mock datasets and uncertainties for the power spectra that could be observed with the DESI and 4HS redshift and peculiar velocity surveys, a hypothetical LSST J<19 based supernova peculiar velocity sample, and HI intensity maps from SKA Phase 1 and Phase 2 Band-2 observations.

We then applied the quadratic model of the power spectrum from ref.~\citep{Bahr_Kalus_2023}
to constrain the turnover scales in these mock datasets and subsequently constrain \(\Omega_m\), \(r_H\) and \(H_0\). We augment this approach by extending the quadratic model to simultaneously fit the velocity or HI temperature power spectra and their cross-power spectra with the galaxy density field. We follow ref.~\citep{Bahr_Kalus_2023} in using mode de-projection to inflate the error bars on the small scale to account for potential systematics, such as the BAO, in all 3 sets of spectra.

We validated our approach by demonstrating that our mock datasets and fitting pipelines yield constraints comparable to those reported in ref.~\citep{Bahr_Kalus_2023}. Our relative uncertainty in the constraint on the turnover scale is approximately 20\% tighter because we do not account for the full survey window function, and the off-diagonal covariance matrices are neglected. Nonetheless, we expect our results to provide a reasonable forecast of the turnover scale constraints for future surveys, particularly regarding the relative benefits of adopting a multi-tracer analysis and combined low- and high-redshift measurements. 

In summary, we find that combining peculiar velocities with galaxy density does not improve the constraints on the turnover scale because for any current or planned sample considered here, either the cosmological volume or the error-weighted number density of peculiar velocities is too low compared to the galaxy density field. However, combining the HI power spectrum with the galaxy power spectrum can significantly improve the detection probability of the turnover and the turnover scale. 4HS and SKA1-B2 have similar constraints on the turnover scale alone. Combining these two surveys improves the turnover scale constraints by \(\sim30\%\) compared to using only the single tracer. The tightest constraint on the turnover scale at low redshift could be achieved by combining 4HS, DESI BGS, and SKA1-B2. This combination gives a \(\sim5\%\) level constraint on the turnover scale and \(\sim 90\%\) detection probability of the turnover. Since these surveys are either ongoing or will commence in the next few years, this goal is achievable within the next decade. We also consider a hypothetical DESI++ catalogue with 31000 \(\mathrm{deg}^2\) sky area and DESI BGS number density. This number density could be achieved by DESI in the northern hemisphere. In the southern hemisphere, this could be achieved by extending the planned 4MOST surveys, or by the Widefield Spectroscopic Telescope in the future. Combining the hypothetical DESI++ catalogue with the SKA2-B2 survey could achieve a \(3\%\) level constraint on the turnover scale and \(\sim97\%\) detection probability of the turnover. 

Furthermore, we demonstrate that in this work, even just combining the DESI BGS with the high redshift LRG, ELG, QSO sample, we could improve the constraint on \(r_H\) and \(\Omega_m\) by \(25\%\) and \(46\%\), respectively, compared to only combining the high redshift LRG, ELG, and QSO samples. This is because the degeneracy direction between \(r_H\) and \(\Omega_m\) changes with the effective redshift. Therefore, adding the low-redshift sample can help break the degeneracy. Replacing DESI BGS with the combination of DESI BGS, 4HS redshifts, and SKA1-B2 could further improve the constraint on \(\Omega_m\) by \(30\%\) and \(r_H\) by 20\%. Finally, the combination of the DESI++ catalogue and SKA2-B2 can further improve this by 14\% and 27\% for \(\Omega_m\) and \(r_H\), respectively. We do caution, however, that despite combining multiple redshift samples, improving the constraints on \(\Omega_m\) and \(H_0\), we still would need to combine with other probes to break the degeneracy between these parameters. 

Overall, our results demonstrate that there is clear potential in constraining the turnover scale at $z<0.5$, and that this can be best achieved through a multi-tracer analysis. Since the HI power spectrum can also be measured at higher redshifts, for example, with SKA-Band 1. Therefore, a multi-tracer approach at high redshift with DESI and SKA Band 1 would be a natural extension of this paper. However, at high redshifts, we do not expect the benefits of a multi-tracer approach to be as apparent — both the DESI and SKA data individually will be quite constraining, and sample variance will not necessarily be the dominant error contribution to such measurements. Hence, we will leave it for future work to investigate applying the multi-tracer approach at higher redshifts. 

\appendix
\section{Appendix}
\subsection{Derivation of the analytical covariance matrix}
\label{sec:anal_cov}
This appendix describes our derivation of the covariance matrix for multi-tracer analyses. Ref.~\citep{Blake_2019} provide the derivation of the auto-covariance matrices (\(C^{gggg}, C^{TTTT}, C^{gTgT}\)) for the galaxy density field and the temperature field. Following section 3.2 of ref.~\citep{Blake_2019} and using Wick's Theorem for multivariate Gaussian distributions, we can deduce that the general form of the covariance matrix is given by 
\begin{equation}
    C^{abcd}(\boldsymbol{k_1}, \boldsymbol{k_2}) = \frac{(2\pi)^3V^2}{2V_{k}}\delta^D(\boldsymbol{k_1}-\boldsymbol{k_2}) \frac{\int d^3 \boldsymbol{x} w_aw_bw_cw_d \left[ \mathcal{Q}_{ac}\mathcal{Q}_{bd} + \mathcal{Q}_{ad}\mathcal{Q}_{bc}\right]}{\mathcal{I}_{cd}\mathcal{I}_{cd}}, 
    \label{eq:cov_blake}
\end{equation}
where 
\begin{equation}
    \mathcal{Q}_{\alpha\beta}(\boldsymbol{k}, \boldsymbol{x}) = \frac{P_{\alpha\beta}(\boldsymbol{k})}{V} \langle f_\alpha(\boldsymbol{x})\rangle \langle f_\beta(\boldsymbol{x})\rangle + \sigma_\alpha(\boldsymbol{x})\sigma_\beta(\boldsymbol{x}) \delta_\alpha^\beta.
    \label{eq:Q_aa}
\end{equation}
The normalisation is given by 
\begin{equation} 
    \mathcal{I}_{\alpha\beta} = \int d^3\boldsymbol{x} w_\alpha(\boldsymbol{x}) w_\beta(\boldsymbol{x}) \langle f_\alpha(\boldsymbol{x})\rangle\langle f_\beta(\boldsymbol{x})\rangle,
\end{equation}
where the weight is \citep{Blake_2019}
\begin{equation}
    w_\alpha(\boldsymbol{k}, \boldsymbol{x}) = \frac{\langle f_\alpha(\boldsymbol{x})\rangle}{\frac{P_{\alpha\alpha}(\boldsymbol{k})}{V}\langle f_\alpha(\boldsymbol{x})\rangle^2 + \sigma_\alpha^2(\boldsymbol{x})}. 
    \label{eq: weight_blake}
\end{equation}
Our results in the main text follow from applying these equations to the appropriate tracer and making some simplifications.

From ref.~\citep{Blake_2019}, we already have starting expressions for the galaxy density field, 
\begin{equation}
    \langle f_g(\boldsymbol{x})\rangle = V\langle \bar{n}_g(\boldsymbol{x})\rangle; \quad \quad \sigma_g^2(\boldsymbol{x}) = V\langle \bar{n}_g(\boldsymbol{x})\rangle, 
    \label{eq: f_g}
\end{equation}
and for the temperature field, 
\begin{equation}
    \langle f_T(\boldsymbol{x})\rangle = \bar{T}_{HI}(\boldsymbol{x}); \quad \quad     \sigma_T^2(\boldsymbol{x}) = \varsigma_T^2(\boldsymbol{x}).
    \label{eq: f_T}
\end{equation}
Comparing equation~(\ref{eq: f_g}) and (\ref{eq: f_T}) definition for \(\langle f(\boldsymbol{x})\rangle\) of temperature and density field, we can define the effective number density for the temperature field as 
\begin{equation}
    \langle \bar{n}_T(\boldsymbol{x})\rangle = \frac{\bar{T}_{HI}(\boldsymbol{x})}{V}.
\end{equation}
Ref.~\citep{Blake_2019} does not give the explicit expression for \(\sigma_T\), but it defines the total noise variance as 
\begin{equation}
    \sigma^2(\boldsymbol{x}) = \frac{1}{V} \sum_i \sigma^2_i\Delta V_i \delta_i,
    \label{eq:sigma_blake}
\end{equation}
where \(\delta_i\) is either one or zero and \(\Delta V_i\) is the volume of the cell \(i\) and \(\sigma_i\) is the noise at cell \(i\). Here, \(\sigma_i\) is similar to the thermal noise per voxel (observed data cube), which in ref.~\citep{Bull_2015} is given by
\begin{equation}
    \sigma_T^2(z) = \frac{T_{\mathrm{sys}}(z)}{\sqrt{n_{\mathrm{pol}}\delta vt_{\mathrm{tot}}}} \frac{\lambda^2}{\theta_B^2 A_e} \sqrt{\frac{\Omega_{\mathrm{sky}}}{\theta_B^2}}\sqrt{\frac{1}{N_dN_b}},
    \label{eq:sigma_T_bull}
\end{equation}
where \(\lambda\) is the wavelength, \(\delta v\) is the resolution of the detector, \(A_e\) is the effective area of the single dish, and \(\theta_B\) is the beam FWHM (Full-Width at Half-Maximum). The corresponding noise covariance matrix is given by \citep{Bull_2015, Santos_2015}
\begin{equation}
    C^N(z) = \frac{\sigma_T^2(z) V_{\mathrm{pix}}}{D_c^2 \frac{c(1+z)^2}{H(z)}}, 
    \label{eq:C_noise}
\end{equation}
where \(V_{\mathrm{pix}}\) is the volume per voxel. To make the link between equation~(\ref{eq:C_noise}) and equation~(\ref{eq:sigma_blake}), which is required for us to use our form of the total covariance matrix, we can simply perform a dimensional/logical comparison. The denominators of both equations are volume, and the numerators of both equations are the noise in each cell multiplied by the volume per cell. Therefore, we can define 
\begin{align}
    \varsigma_T^2(\boldsymbol{x}) = C^N(z) &= \frac{T_{\mathrm{sys}}^2(z)\Omega_{\mathrm{sky}}\Delta\tilde{\nu}}{n_{\mathrm{pol}}t_{\mathrm{tot}} \Delta \nu} \left(\frac{\lambda^4}{A_e^2\theta_B^4}\right) \frac{1}{N_dN_b} \nonumber \\
    &=\frac{T_{\mathrm{sys}}^2(z)\Omega_{\mathrm{sky}}}{n_{\mathrm{pol}}t_{\mathrm{tot}} \nu_{21}} \left(\frac{4}{\eta\pi}\right)^2 \frac{1}{N_dN_b}
    \label{eq:sigmaT_new}
\end{align}
consistent with equation~(\ref{eq:sigma_TT}). The noise covariance matrix only depends on the redshift, as we have ignored the effect of instrumental beams. To simplify equation~(\ref{eq:sigmaT_new}) to the second line, we use \(\nu_{21} = \frac{\Delta \nu}{\Delta \tilde{\nu}}\), \(A_e = \eta\pi\left(\frac{D_{\mathrm{dish}}}{2}\right)^2\), and \(\theta_B \approx \frac{\lambda}{D_{\mathrm{dish}}}\). Here, \(D_{\mathrm{dish}}\) is the diameter of a single dish and \(\Delta \nu\) is the frequency range of the detector. In Appendix~\ref{sec:HI_noise}, we demonstrate that our covariance matrix with equation~(\ref{eq:sigmaT_new}) as the noise model agrees well with the covariance matrix from ref.~\citep{Bull_2015} which can be computed with the \textsc{bao21cm} software provided in that reference.  

For the velocity field, ref.~\citep{Howlett_2019} derives the auto-covariance matrices. Based on their definition of the weight for the velocity field 
\begin{equation}
    w_v(\boldsymbol{k}, \boldsymbol{x}) = \frac{1}{\Sigma_v(\boldsymbol{x}) + \bar{n}_v(\boldsymbol{x}) P_{vv}(\boldsymbol{k})}.
    \label{eq: w_v_howlett}
\end{equation}
Comparing equation~(\ref{eq: w_v_howlett}) to equation~(\ref{eq: weight_blake}), we get 
\begin{equation}
    \langle f_v(\boldsymbol{x}) \rangle = V\bar{n}_v(\boldsymbol{x}); \quad \quad \sigma_v(\boldsymbol{x}) = \sqrt{\Sigma_v(\boldsymbol{x}) V\bar{n}_v(\boldsymbol{x})}, 
    \label{eq: f_v}
\end{equation}
where we need the extra factor of \(V\) to cancel out the volume factor in equation~(\ref{eq:cov_blake}) compared to how the covariance matrix is defined in ref.~\citep{Howlett_2019}.

Our next steps are to modify equation~(\ref{eq:cov_blake}) to accommodate different binned power spectrum multipoles and rewrite the volume integral as an integral over redshift. Based on ref.~\citep{Grieb_2016}, the general covariance matrix for the power spectrum multipoles is given by 
\begin{align}
    C^{abcd}_{l_1l_2}(k_1, k_2) &= \frac{(2\pi)^3V^2}{2V_{k}}\delta^D(k_1-k_2)(2l_1 + 1)(2l_2+1) \Omega_{\mathrm{tracer}}\int d\mu \mathcal{L}_{l_1}(\mu_1) \mathcal{L}_{l_2}(\mu_2) \nonumber \\
    &\frac{\int dz D_c^2 \frac{dD_c}{dz} w_aw_bw_cw_d \left[ \mathcal{Q}_{ac}\mathcal{Q}_{bd} + \mathcal{Q}_{ad}\mathcal{Q}_{bc}\right]}{\mathcal{I}_{cd}\mathcal{I}_{cd}},
    \label{eq:C_l1_l2}
\end{align}
where the normalisation term is now
\begin{equation} 
    \mathcal{I}_{\alpha\beta}(z, k, \mu) = \Omega_{\mathrm{tracer}}\int dz D_c^2(z) \frac{dD_c}{dz} w_\alpha(z, k, \mu) w_\beta(z, k, \mu) \langle f_\alpha(z, \mu)\rangle\langle f_\beta(z, \mu)\rangle,
    \label{eq: I_ab_l},
\end{equation}
and the kernel \(\mathcal{Q}_{\alpha \beta}\) is given by 
\begin{equation}
    \mathcal{Q}_{\alpha \beta}(z, k, \mu) = \frac{P_{\alpha\beta}(z, k, \mu)}{V} \langle f_\alpha(z, \mu)\rangle\langle f_\beta(z, \mu)\rangle + \sigma_{\alpha}(z, \mu)\sigma_{\beta}(z, \mu)\delta_\alpha^\beta,
\end{equation}
with
\begin{equation}
    w_\alpha(z, k, \mu) = \frac{\langle f_\alpha(z, \mu)\rangle}{\frac{P_{\alpha\alpha}(z, k, \mu)}{V}\langle f_\alpha(z, \mu)\rangle^2 + \sigma_\alpha^2(z, \mu)}. 
    \label{eq: weight_blake_new}
\end{equation}

Given these, the equations in the main text can be derived by simply substituting in the various expressions for $f_\alpha$ and $\sigma_\alpha$ given above, canceling out the various factors of volume $V$ that arise in the numerator and denominator, assuming that our weights depend only on redshift (such that the normalization constants can be moved outside the integrals) and defining both the effective number density for intensity mapping as given in equation~(\ref{eq: n_T}) and

%By taking the \(\frac{1}{V}\) term in \(\mathcal{Q}^{\alpha\beta}\) outside the integral and substitute in the definition for \(N_m\), equation~(\ref{eq:C_l1_l2}) becomes
%\begin{align}
%    C^{abcd}_{l_1l_2}(k_1, k_2) &= \frac{(2\pi)^3}{2V_k}\delta^D(k_1-k_2)(2l_1 + 1)(2l_2+1) \Omega_{\mathrm{tracer}}\int d\mu \mathcal{L}_{l_1}(\mu_1) \mathcal{L}_{l_2}(\mu_2) \nonumber \\
%    &\frac{V^4\int dz D_c^2 \frac{dD_c}{dz} w_aw_bw_cw_d \left[ \mathcal{R}^{ac}\mathcal{R}^{bd} + \mathcal{R}^{ad}\mathcal{R}^{bc}\right]}{\mathcal{I}_{cd}\mathcal{I}_{cd}},
%    \label{eq: C_abcd_l1_l2_new}
%\end{align}
%where  
%\begin{equation}
%   \mathcal{R}^{\alpha \beta}(z, k, \mu) = \bar{n}_{\alpha}(z, \mu)\bar{n}_{\beta}(z, \mu) P_{\alpha\beta}(z, k, \mu) +  \frac{\sigma_{\alpha}(z, \mu)\sigma_{\beta}(z, \mu)}{V}\delta_\alpha^\beta
%\end{equation}
%otherwise. Here, we define \(\bar{n}_\alpha = \frac{\langle f_\alpha(z, \mu)\rangle}{V}\). For galaxy density and peculiar velocity, this corresponds to the usual definition of number density based on equations~(\ref{eq: f_g}) and (\ref{eq: f_v}). The effective number density for intensity mapping is given in equation~(\ref{eq: n_T}). We can also conveniently define 
\begin{equation}
    \Sigma_\alpha = \frac{\sigma^2_\alpha(z, \mu)}{\langle f_\alpha(z, \mu)\rangle}.
\end{equation}
%Then, \(\frac{\sigma^2_{\alpha}(z, \mu)}{V} = \bar{n}_\alpha(z, \mu) \Sigma_\alpha(z, \mu)\). Substituting in the definition for \(\Sigma_\alpha\) and \(\bar{n}_\alpha\), the normalisation factors in the denominator becomes 
%\begin{equation}
%    \mathcal{I}_{\alpha\beta} = \Omega_{\mathrm{tracer}}V^2\int dz D_c^2(z) \frac{dD_c}{dz} w^{l_1}_\alpha(z, \mu) w^{l_2}_\beta(z, \mu) \bar{n}_\alpha(z, \mu) \bar{n}_\beta(z, \mu).
%    \label{eq:I_ab_l_new}
%\end{equation}
%Since the denominator of equation~(\ref{eq: C_abcd_l1_l2_new}) is the product of two normalisation factors, we have a factor of \(V^4\) which cancels out the \(V^4\) factor in the numerator. This recovers the covariance matrix formula in equation~(\ref{eq:cov_full}).

%For the optimal weight, we can rearrange equation~(\ref{eq: weight_blake_new}) to get 
%\begin{equation}
%    w_\alpha(z, k, \mu) = \frac{V}{\langle f_\alpha(z, \mu)\rangle \left(P_{\alpha\alpha}(z, k, \mu) + \frac{\sigma^2_\alpha(z, \mu) V}{\langle f_\alpha(z, \mu)\rangle^2}\right)}.  
%\end{equation}
%If we substitute in the definition for \(\bar{n}_\alpha\) and \(\Sigma_\alpha\), we will recover equation~(\ref{eq:FKP}). 

To define the effective redshift for our various tracers, we can take the $l_{1}=l_{2}=0$ multipole moments for the auto-covariance in~\ref{eq:cov_full}. %, the auto-covariance matrix of the auto-power spectrum is given by 
%\begin{align}
%    C^{aaaa}_{l_1l_2}(k,k^{\prime}) &= \frac{(2\pi)^3(2l_1+1)(2l_2+1)}{V_k(I_{aa})^2}\delta^{D}(k-k^{\prime}) \Omega_{\mathrm{tracer}}\int_{-1}^{1}d\mu \mathcal{L}_{l_1}(\mu) \mathcal{L}_{l_2}(\mu) \int dz\, D_{c}^{2}(z)\frac{dD_{c}}{dz} \nonumber \\ &w_a^4(z) \biggl[\mathcal{P}^{aa}(k,\mu,z)\biggl]^2.  
%    \label{eq:cov_auto}
%\end{align}
We can then follow ref.~\citep{Tegmark_1997} by assuming the power spectrum is independent of redshift to obtain a covariance matrix that can be written as \(C \propto \frac{P(k)^2}{V_k V_{\mathrm{eff}}}\). We do this by substituting in the expressions for the normalisation and optimal weights and simplifying, in which case
\begin{equation}
    C^{aaaa} \propto \frac{1}{V_k\Omega_{\mathrm{tracer}} \int dz D_c^2 \frac{dD_c}{dz} \left(\frac{1}{P_{aa}(k) + \frac{\Sigma_a(z)}{\bar{n}_a(z)}}\right)^2}. 
\end{equation}

We can multiply both the numerator and denominator by \((P_{aa}(k))^2\), to define the effective volume as 
\begin{equation}
    V_{\mathrm{eff}}^a(k) = \Omega_{\mathrm{tracer}} \int dz D_c^2 \frac{dD_c}{dz} \left(\frac{P_{aa}(k)}{P_{aa}(k) + \frac{\Sigma_{a}(z, \mu)}{\bar{n}_{a}(z)}}\right)^2,
\end{equation}
which matches the result in the main text. Substituting in the definition of \(\Sigma_g\) for the galaxy power spectrum, we recover the effective volume formula in ref.~\citep{Tegmark_1997}. 

\subsection{HI noise covariance matrix comparison}
\label{sec:HI_noise}
In order to test our analytic covariance matrix for the HI intensity mapping surveys, we also make use of the \textsc{bao21cm} code from ref.~\citep{Bull_2015} and compare the two. We use the survey parameters for SKA1-B2 from Table 2 of ref.~\citep{Bull_2015} for this comparison to match the \textsc{bao21cm} code. This ensures the covariance matrices are compared on an equal footing. In this section, we compare the relative uncertainty of the power spectrum \(\Delta P /P\) and the turnover constraints from these two different codes. 

\begin{figure}
    \centering
	\includegraphics[width=0.70\textwidth]{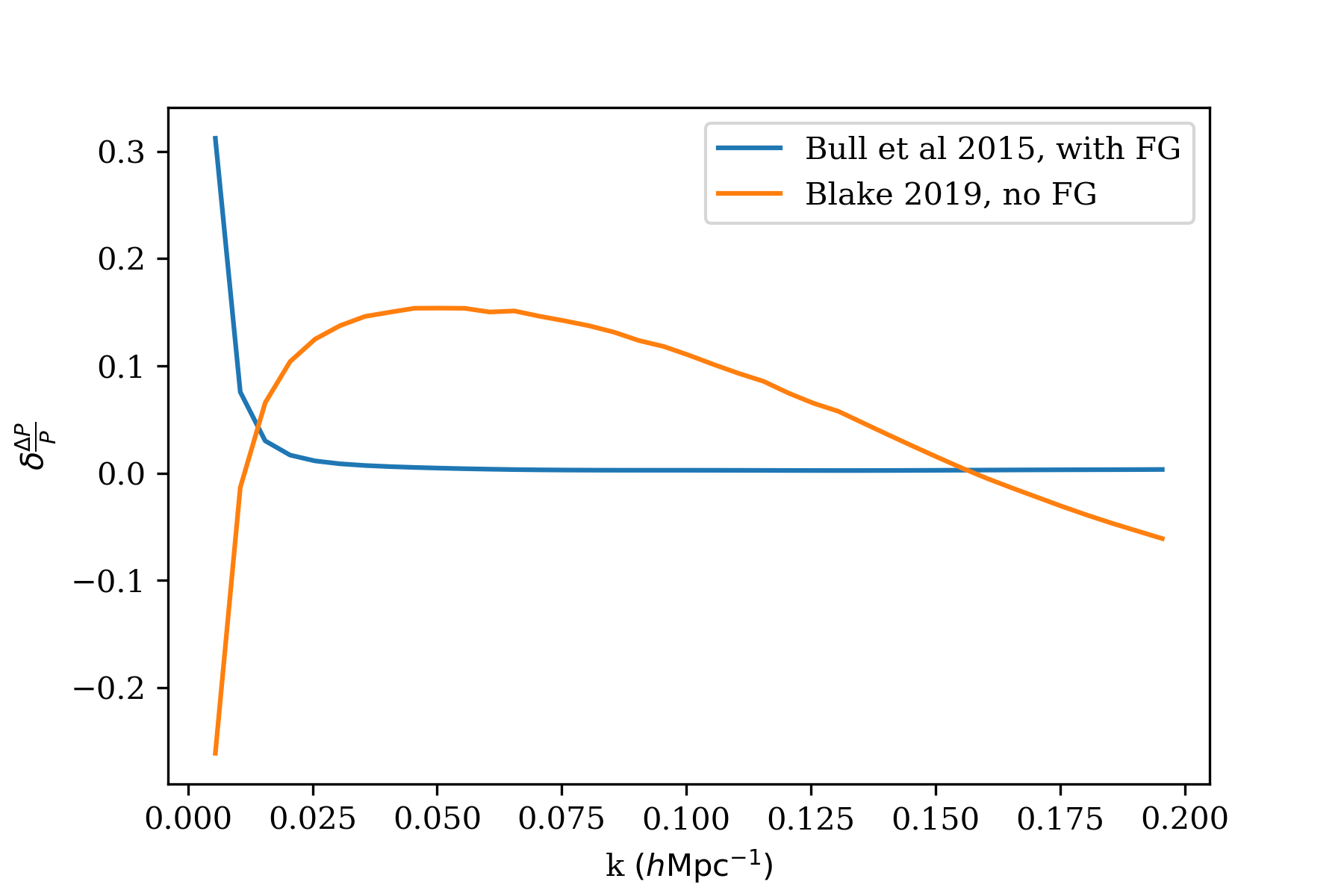}
    
    \caption{The relative difference between our model (orange; based on \cite{Blake_2019}), ref.~\citep{Bull_2015} with residual foreground (red), and the fiducial model (ref.~\citep{Bull_2015} model with perfect foreground (FG) subtraction). Our model agrees within 20\% with the perfect background subtraction model from ref.~\citep{Bull_2015}. The residual foreground mainly affects the uncertainty of the power spectrum on the large scales.}
    \label{fig:delta_P_bull}
\end{figure}

\begin{figure}
    \centering
	\includegraphics[width=0.70\textwidth]{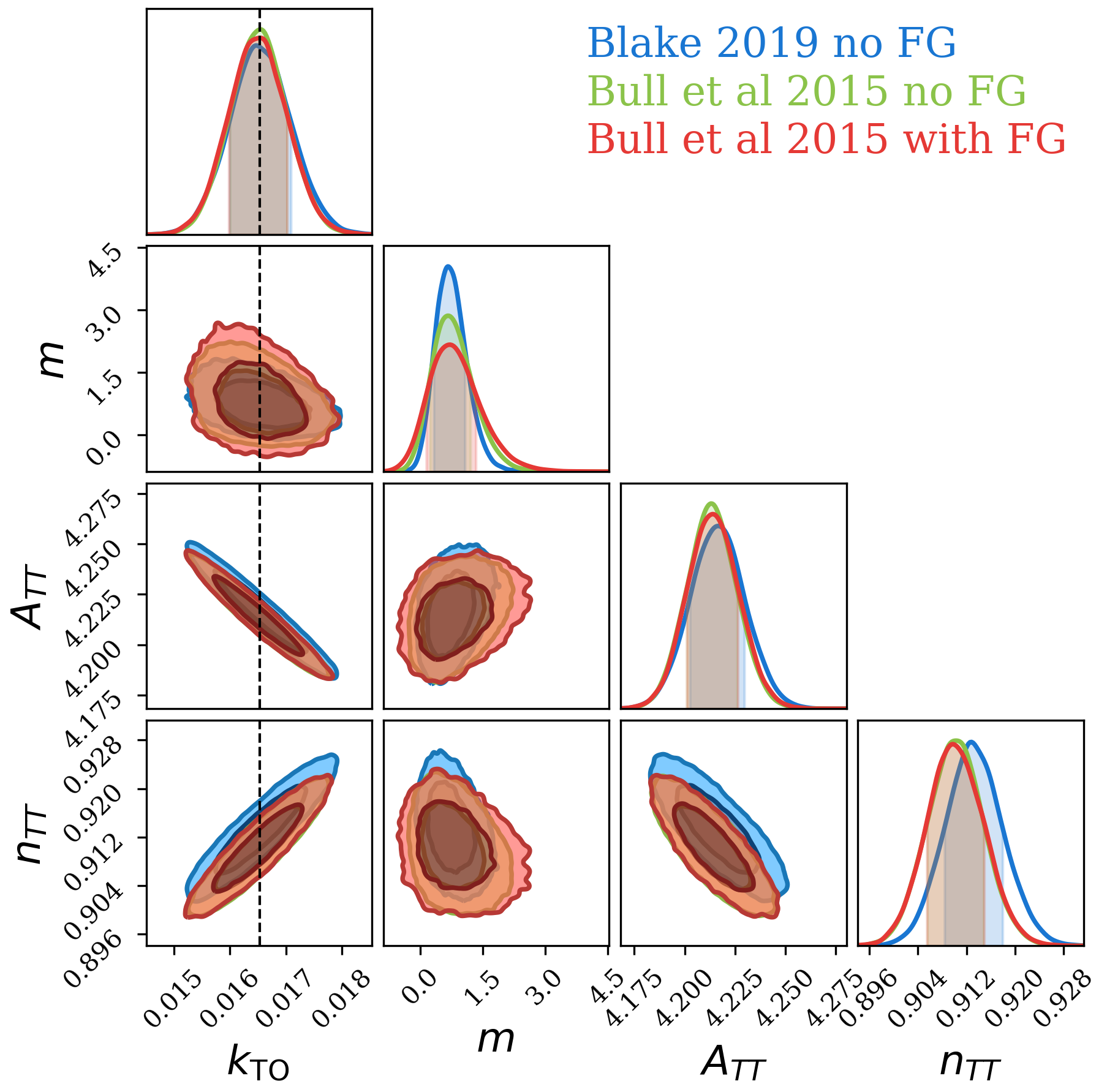}
    
    \caption{The constraints on the turnover scale with the three covariance matrices in Fig.~\ref{fig:delta_P_bull}. All three covariance matrices give similar constraints on the turnover scales and the detection probability of the turnover.}
    \label{fig:FKP_vs_Bull}
\end{figure}

Fig.~\ref{fig:delta_P_bull} illustrates the relative difference between our model (based on \cite{Blake_2019}), the model of ref.~\citep{Bull_2015} with residual foreground, and a fiducial method (ref.~\citep{Bull_2015} model with perfect foreground subtraction). The relative difference is given by 
\begin{equation}
    \delta_i \frac{\Delta P}{P} = \frac{\frac{\Delta P_i}{P_i} - \frac{\Delta P_{\mathrm{fid}}}{P_{\mathrm{fid}}}}{\frac{\Delta P_{\mathrm{fid}}}{P_{\mathrm{fid}}}},
\end{equation}
where ``fid" denotes the fiducial model. We use \textsc{bao21cm}'s default foreground subtraction residual amplitude \(\epsilon_{\mathrm{FG}} = 10^{-6}\) to compute the blue curve \citep{Bull_2015}. The residual noise mainly affects the uncertainty of the power spectrum on large scales. Our model is consistent within 20\% of the \textsc{bao21cm} model. Unlike \textsc{bao21cm}, our code does not account for the effect of instrumental beams. However, ref.~\citep{Blake_2019} demonstrates that such effects mainly affect the power spectrum on small scales. Therefore, we do not expect it to affect the turnover scale constraints as illustrated in Fig.~\ref{fig:FKP_vs_Bull}. The difference in constraints on the turnover scale between the fiducial method and our code is around 6.4\%. However, the detection probability from our code is approximately 2.1\% higher than that of the fiducial method. These differences are small, so ignoring the effect of instrumental beams does not significantly impact our main results. Furthermore, the constraint on the turnover scale is only approximately 2\% weaker, and the detection probability is reduced by approximately 3\% when including the residual foreground. This also shows that the residual foreground will have little impact on the main results, 

\subsection{The effect of different models of the HI power spectrum}\label{sec:HI_model}
\begin{figure}
    \centering
	\includegraphics[width=0.70\textwidth]{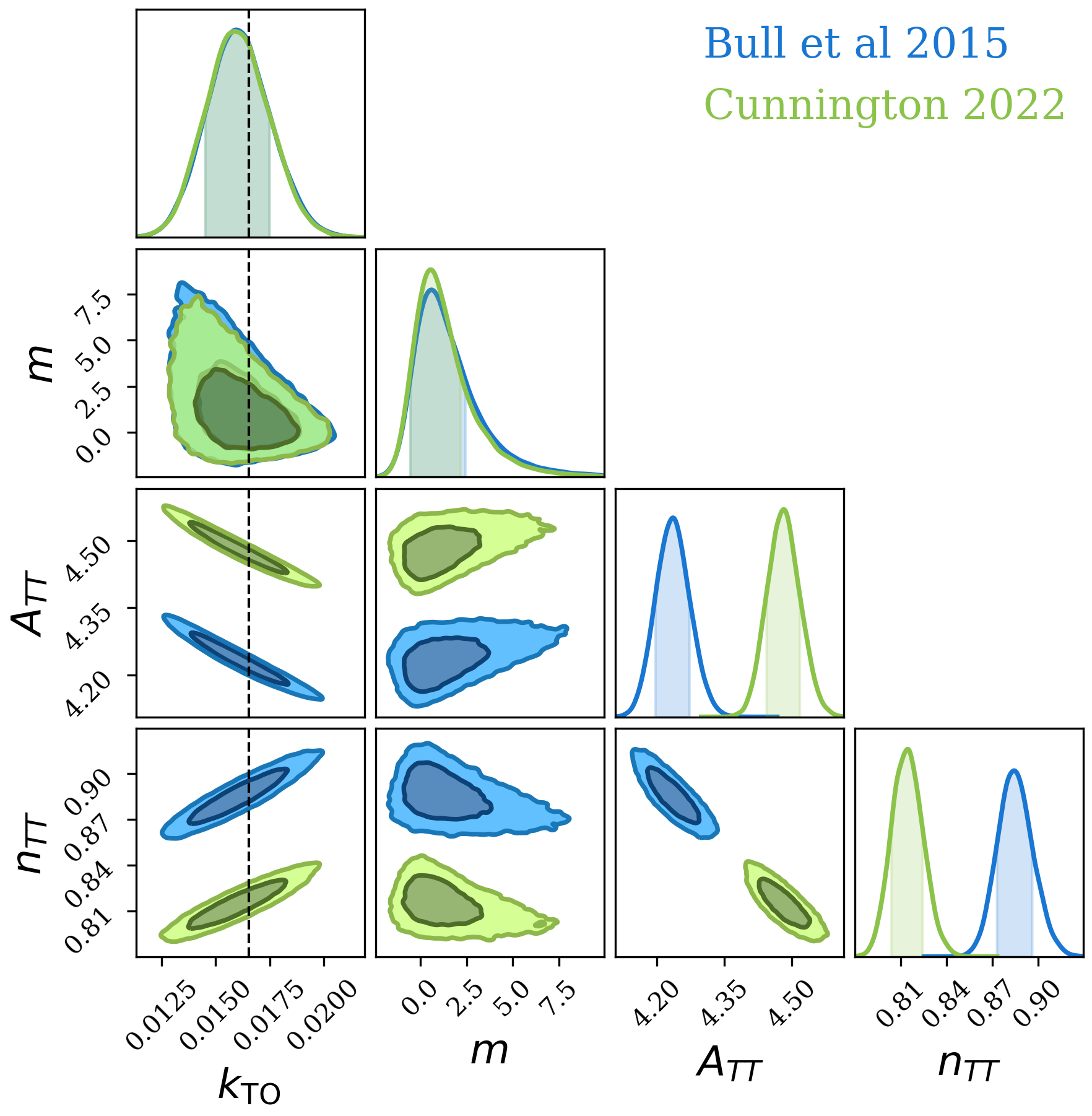}
    
    \caption{The constraints on the turnover scales from SKA1-B2 using the HI power spectrum model adopted in this work (blue) and ref.~\citep{Cunnington_2022} (green). Both return consistent constraints on the turnover scale and turnover detection probability. The differences between the two models have been mitigated by the slope parameter \(n_{TT}\) and the amplitude parameter \(A_{TT}\).} 
    \label{fig:Cunnington_vs_Bull}
\end{figure}

Ref.~\citep{Cunnington_2022} forecast the constraints on the turnover scale with the SKA Phase 1 Band 1 survey. However, they use a different model for the redshift evolution of the HI bias \citep{Villaescusa-Navarro_2015}, the HI mean temperature \citep{Battye_2013}, and the HI energy density \citep{Pourtsidou_2017, SKA_2020}
\begin{align}
    b_{HI}(z) &= 0.842 + 0.693z-0.046z^2 \\
    \bar{T}_{HI}(z) &= 180 \Omega_{HI}(z)h\frac{(1+z)^2H_0}{H(z)} \\
    \Omega_{HI}(z) &= 0.00048 + 0.00039 - 0.000065z^2.
\end{align}
Fig.~\ref{fig:Cunnington_vs_Bull} illustrates that the differences between our parameterisation and that above are mitigated by the slope parameter \(n_{TT}\) and the amplitude parameter \(A_{TT}\). The turnover scale constraints and the turnover detection probability remained unchanged. 

\acknowledgments
We thank David Parkinson and Benedict Bahr-Kalus for their insightful discussion on their model of the turnover scale. YL, CH, and TD acknowledge support from the Australian Government through the Australian Research Council’s Laureate Fellowship (project FL180100168) and Discovery Project (project DP20220101395) funding schemes. YL is also supported by an Australian Government Research Training Program Scholarship.

\bibliographystyle{JHEP}
\bibliography{example.bib}


\end{document}